\documentclass[12pt,a4paper]{article}
\usepackage{amsmath,amssymb,mathrsfs,framed,esint,slashed,upgreek,graphicx}
\usepackage[colorlinks]{hyperref}
\usepackage{color}
\usepackage[all]{xy}

\usepackage{soul}

\newlength{\bibitemsep}
\newlength{\bibparskip}
\let\oldthebibliography\thebibliography
\renewcommand\thebibliography[1]{%
  \oldthebibliography{#1}%
  \setlength{\parskip}{\bibitemsep}%
  \setlength{\itemsep}{\bibparskip}%
}

\DeclareMathAlphabet{\mathpzc}{OT1}{pzc}{m}{it}

\newcommand{\rem}[1]{}

\newcommand{\de}{{\rm d}}

\newcommand{\bA}{{\boldsymbol{\cal A}}}

\newcommand{\beq}{\begin{equation}}
\newcommand{\eeq}{\end{equation}}
\newcommand{\ben}{\begin{eqnarray}}
\newcommand{\een}{\end{eqnarray}}

\numberwithin{equation}{section}

\begin{document}

\title{\vspace{-1.3cm}Hybrid quantum-classical dynamics\\of pure-dephasing systems\footnote{Contribution to the Special Collection \href{https://iopscience.iop.org/journal/1751-8121/page/koopman-methods-in-classical-and-quantum-classical-mechanics}{``Koopman methods in classical and quantum-classical mechanics''} in the Journal of Physics A.}\vspace{-.2cm}
}
\author{Giovanni Manfredi$^1$, Antoine Rittaud$^1$, Cesare Tronci$^{2,3}$ \smallskip\vspace{-.1cm}
\\
\footnotesize
\it\hspace{-1.17cm} $^1$Universit\'e de Strasbourg, CNRS, Institut de Physique
 et Chimie des Mat\'eriaux de Strasbourg, 67000 Strasbourg, France
\\
\footnotesize
\it\hspace{-1.17cm} $^2$Department of Mathematics, University of Surrey, Guildford GU2 7XH, United Kingdom
\\
\footnotesize
\it\hspace{-1.17cm} 
$^3$Department of Physics and Engineering Physics, Tulane University, New Orleans, LA 70118, United States}
\date{\vspace{-1.3cm}}

\maketitle

\begin{abstract}
We consider the interaction dynamics of a classical oscillator and a quantum two-level  system for different pure-dephasing Hamiltonians of  the type  $\widehat{H}(q,p)=H_C(q,p)\boldsymbol{1}+H_I(q,p)\widehat\sigma_z$. This type of systems represents a severe challenge for popular hybrid quantum-classical descriptions. For example, in the case of the common Ehrenfest model, the classical density evolution is shown to decouple entirely from the pure-dephasing quantum dynamics. We focus on a recently proposed hybrid wave equation that is based on  Koopman's wavefunction description of classical mechanics. This model retains quantum-classical correlations  whenever a coupling potential is present.  Here, several benchmark problems are considered and the results are compared with those arising from fully  quantum  dynamics.  A good agreement is  found for a series of study cases involving harmonic oscillators with linear and quadratic coupling, as well as time-varying coupling parameters. In all these cases the classical evolution coincides exactly with the oscillator dynamics resulting from the fully quantum description. In the special case of time-independent coupling involving a classical oscillator with varying frequency, the  quantum Bloch rotation exhibits peculiar features that escape from the hybrid description. In addition, nonlinear corrections to the harmonic Hamiltonian lead to an overall growth of decoherence at long times, which is absent in the fully quantum treatment.
\end{abstract}

{\vspace{-1.3cm}
\small
\tableofcontents
}
\addtocontents{toc}{\protect\setcounter{tocdepth}{3}}

\section{Introduction}

The search for models describing the interaction dynamics of quantum and classical systems goes back to the early days in the history of quantum mechanics and is typically related to the measurement problem. In that context, a quantum system is  coupled to a classical system comprising a pointer and the environment, that is the rest of the measuring apparatus. While the environment is usually thought of as a heat bath  inducing irreversible thermodynamical processes, the system-pointer interaction is an intrinsically unitary process known as \emph{pre-measurement} \cite{Peres,Sudarshan}.

Besides its relevance in such foundational issues, the quantum-classical coupling problem has also emerged over the years in a series of different contexts. In theoretical physics, for example, the possibility that the theory of gravity may not be quantized leads to consider  coupling  the  classical Einstein tensor  to quantum matter \cite{carlip}. In this context, the theory of semiclassical gravity is based on a mean-field model that fails to retain quantum-classical correlations \cite{boucher}.
Furthermore, several activities on quantum-classical coupling are currently motivated by the necessity of mitigating the computational costs of fully quantum many-body simulations. For example, in chemical physics molecular dynamics algorithms often approximate nuclei as classical while retaining the full quantum effects of electron dynamics \cite{CrBa18,Kapral,Tully}. In solid state physics, recent proposals \cite{HuHeMa17} suggest treating the orbital degrees of freedom as classical while leaving spins as fully quantum observables. Similar quantum-classical descriptions have also appeared in spintronics, where quantum spins are coupled to classical ferromagnets thereby reaching unexplored parameter regimes in quantum control \cite{RuKaUp22}.

\subsection{Models of hybrid quantum-classical dynamics}
Despite several attempts \cite{Aleksandrov,boucher,Diosi,Gerasimenko,JaSu10,PrKi,WiSr92}, most hybrid quantum-classical models suffer from various consistency issues \cite{AgCi07,boucher,Ghose,PeTe,Salcedo,Terno}. For example, some models allow for the violation of the uncertainty principle, while some others fail to return uncoupled quantum and classical motion in the absence of an interaction potential. Following \cite{boucher}, we can propose five fundamental properties that quantum-classical dynamical models should possess \cite{GBTr21}:
 \begin{enumerate}
\item the classical system is identified by a positive probability density on phase space at all times;
 \item the quantum system is identified by a positive-semidefinite density operator $\hat\rho$ at all times;
 \item the model is covariant under both quantum unitary transformations and classical canonical transformations;
 \item  in the absence of an interaction potential, the model reduces to uncoupled quantum and classical dynamics;
 \item in the presence of an interaction potential, the {\it quantum purity} $\operatorname{Tr}\hat\rho^2$ is not a constant of motion (decoherence property).
 \end{enumerate}
Notice that the last property rules out the standard mean-field model, which entirely neglects quantum-classical correlations thereby conserving purity. To our knowledge, the only available model satisfying all five properties  is the \emph{Ehrenfest model}:
\beq\label{Ehrenfest}
\frac{\partial D}{\partial t}+\bigg\langle\frac{\partial \widehat{H}}{\partial p}\bigg\rangle\frac{\partial D}{\partial q}-\bigg\langle\frac{\partial \widehat{H}}{\partial q}\bigg\rangle\frac{\partial D}{\partial p}=0
\,,\qquad\quad
\frac{\partial \psi}{\partial t}+\bigg\langle\frac{\partial \widehat{H}}{\partial p}\bigg\rangle\frac{\partial \psi}{\partial q}-\bigg\langle\frac{\partial \widehat{H}}{\partial q}\bigg\rangle\frac{\partial \psi}{\partial p}=-\frac{i}\hbar\widehat{H}\psi.
\quad
\eeq
 Here, $D$ is the classical Liouville distribution, while the quantum density matrix is $\hat\rho=\int\! D\psi\psi^\dagger\,\de q\de p$. Also, we have denoted $\langle\widehat{A}\rangle=\langle\psi,\widehat{A}\psi\rangle$, while the hybrid Hamiltonian $\widehat{H}=\widehat{H}(q,p)$ is a phase-space function with values in the space of Hermitian operators on the quantum Hilbert space. Despite its popularity in chemical physics \cite{Alonso}, the  model \eqref{Ehrenfest} is well known to be unreliable in reproducing the correct decoherence levels arising from the fully quantum description. In particular, the Ehrenfest model suffers from the \emph{overcoherence problem}, that is it typically leads to higher purity levels than those arising from the fully quantum treatment \cite{AkLoPr14}. In some cases, a certain \emph{undercoherence} was also reported  \cite{HuGuFr18}.  As we will see, the Ehrenfest model suffers also from other serious issues that become manifest in the class of Hamiltonians considered in this paper.

Several other hybrid quantum-classical models are available, although they generally fail to fulfill all the five properties listed  above. In some cases this lack of consistency leads to unrealistic timescales of the quantum evolution \cite{BoGBTr19}. A candidate model beyond Ehrenfest which satisfies all the five properties was recently presented in \cite{GBTr23,GBTr22,GBTr21}. In that case, the quantum-classical interaction triggers challenging nonlinear terms and suitable closure schemes are currently under development. This nonlinear model was formulated upon revisiting a \emph{linear} quantum-classical wave equation previously proposed in \cite{BoGBTr19}. This equation governs the unitary evolution of a square-integrable hybrid wavefunction $\Upsilon(q,p,x)$, where $x$ is the quantum position coordinate. In its simplest form, the quantum-classical wave equation reads $\partial_t\Upsilon+\{\Upsilon,\widehat{H}\}=i\hbar^{-1}(p\partial_p\widehat{H}-\widehat{H})\Upsilon$. Satisfying the properties 2)-5) above, this equation was also shown to retain property 1) when the  Hamiltonian depends on the quantum degrees of freedom through a set of mutually commuting observables \cite{GBTr20} (\emph{pure-dephasing Hamiltonian}). However, the hybrid wave equation  is not known to satisfy property 1) in the general case. In addition, while  this equation is covariant under gauge transformations $\Upsilon\mapsto e^{i\varphi(q,p)}\Upsilon$, it is not gauge-invariant. This kind of gauge invariance was indicated as a desirable property  in \cite{boucher}, based on the fact that classical phases  cannot lead to measurable effects \cite{Sudarshan}. Indeed, this gauge invariance plays a key role in the nonlinear model from \cite{GBTr23,GBTr22,GBTr21}.

While these aspects deserve appropriate care, the hybrid wave equation from \cite{BoGBTr19} seems to represent a step forward in the current state of the art of quantum-classical modeling. For example, unlike  widely popular models in chemical physics \cite{Aleksandrov,Gerasimenko,Kapral}, the hybrid wave equation preserves the Heisenberg uncertainty principle (as a consequence of property 2). Also, it reduces to uncoupled quantum and classical mechanics in the absence of an interaction potential (property 4), unlike alternative approaches based on master equations \cite{Diosi} or   quantum hydrodynamics \cite{FoHoTr19,Hall}. As we will see, it also overcomes the Ehrenfest model in terms of its predictions on the  dynamics of the classical system. Last, the hybrid wave equation was recently shown to be Galilean-covariant \cite{Andre}, thereby ensuring another desirable property. Given these aspects of merit, we are motivated to present a benchmark study comparing the results obtained from the quantum-classical wave equation with those arising from fully quantum dynamics. In particular we will consider the simple case of pure-dephasing Hamiltonians, for which the classical density was shown to be always positive-definite \cite{GBTr20}.
Pure-dephasing problems provide a suitable benchmark platform for our proof-of-principle study of the mathematical model under consideration, while  the development of a cost-effective  computational method is left for future work. For example, the presence of characteristic curves in its Madelung form makes the quantum-classical wave equation amenable to trajectory-based methods, as opposed to the more expensive (yet more accurate) finite-volume schemes adopted here.

This paper will study the case of harmonic and anharmonic oscillators coupled linearly and nonlinearly to a quantum  two-level system,  and  focus especially on the  dynamics of the quantum subsystem. In addition, we will  present some results on the quantum-classical spin-momentum correlations and we will also consider the case of time-dependent parameters. While predicting slightly higher levels of decoherence, the hybrid results are in  good  agreement with the quantum dynamics when the hybrid Hamiltonian $\widehat{H}(q,p)$ is a quadratic function of the classical coordinates. One exception is given by  the particular case when the oscillator frequency varies with time. In that case, the fully quantum dynamics displays rather peculiar features that cannot be reproduced by the hybrid wave model. Furthermore, in the case of  non-harmonic  hybrid dynamics, the decoherence  becomes particularly pronounced over time thereby leading to distinctive differences from the fully quantum evolution.

\subsection{Quantum-classical wave equation}
The formulation of the quantum-classical wave equation (QCWE) exploits an earlier suggestion from George Sudarshan \cite{Marmo,Sudarshan}. The idea is to resort to Koopman's wavefunction description of classical dynamics \cite{Jo20,Koopman,VonNeumann2} in such a way that the hybrid quantum-classical motion can be formulated on the tensor product Hilbert space comprising products of classical and quantum wavefunctions. The general theory of the resulting model for hybrid quantum-classical wavefunctions has been presented in several instances \cite{BoGBTr19,GBTr22,GBTr21,GBTr20,TrJo21}. At this stage we simply present its hybrid wave equation in the general form:
\beq\label{QCWE}
\frac{\partial \Upsilon}{\partial t}-\{ \widehat{H}, \Upsilon\}=\frac{i}\hbar\bigg({\cal A}_q\frac{\partial \widehat{H}}{\partial p}-{\cal A}_p\frac{\partial \widehat{H}}{\partial q}-\widehat{H}\bigg)\Upsilon
\,.
\eeq
As mentioned previously, $\Upsilon(q,p,x)$ is a wavefunction depending on the classical phase-space coordinates  $\mathbf{z} \equiv (q,p)$ as well as the quantum coordinate $x$. The right-hand side of \eqref{QCWE} deserves some comments. First, the vector potential $\bA=({\cal A}_q,{\cal A}_p)$ is called \emph{symplectic potential} and its defining relation is
\[
\nabla \bA-(\nabla\bA)^T=- \mathbb{J}
,\quad\qquad\text{ with }\quad\qquad
 \mathbb{J}=\left(\begin{matrix}
0 & 1
\\
-1 & 0
\end{matrix}\right),
\]
 where the gradient $\nabla=(\partial_q, \partial_p)$ is computed in the phase space.
The symplectic potential is defined up to a pure gradient gauge term. In the standard gauge, one has $\bA=(p,0)$ so that in the purely classical case $\widehat{H}=H\boldsymbol{1}$ the parenthesis on the right-hand side of \eqref{QCWE} reduces to the phase-space Lagrangian  $\mathscr{L}=p\partial_p{H}-{H}$. Indeed, this term on the right-hand side of \eqref{QCWE} has the role of incorporating the  phase evolution in the classical sector \cite{deGo04,GBTr20}. Another relevant gauge is given by
\[
\bA=(p,0)-\frac12\nabla(pq)=\frac12(p,-q)\,.
\]
 In this gauge, the QCWE reads
$$\frac{\partial \Upsilon}{\partial t}-\{ \widehat{H}, \Upsilon\}=\frac{i}\hbar\bigg(\frac{p}2\frac{\partial \widehat{H}}{\partial p}+\frac{q}2\frac{\partial \widehat{H}}{\partial q}-\widehat{H}\bigg)\Upsilon ,
$$
which is the form used throughout this paper. This \emph{harmonic gauge} is particularly convenient for purely quadratic Hamiltonians since it makes the right-hand side of \eqref{QCWE} vanish identically.

In the QCWE setting, the quantum density matrix is given by
\beq
\hat{\rho}= \int \Upsilon(q,p)\Upsilon^\dagger(q,p)\,\de q \de p
\,,
\label{qden}
\eeq
which is consistently positive definite thereby ensuring that the uncertainty principle is satisfied at all times. Here, the notation is such that $\Upsilon^\dagger(q,p)$ is the quantum adjoint, so that both $\Upsilon_1^\dagger\Upsilon_2=\langle\Upsilon_1|\Upsilon_2\rangle$ and $\Upsilon^\dagger\Upsilon=\|\Upsilon\|^2$ identify   distributions on the phase space. In addition, the classical distribution reads
\beq
\rho_c(q,p)=\|\Upsilon(q,p)\|^2-\operatorname{div}(\|\Upsilon(q,p)\|^2\Bbb{J}\boldsymbol{\cal A}(q,p))+\hbar\operatorname{Im}\{\Upsilon^\dagger(q,p),\Upsilon(q,p)\}
\,.
\label{cden}
\eeq
While this expression is generally sign-indefinite,  the corresponding distribution  was shown to remain positive for pure-dephasing Hamiltonians, that is  Hamiltonians depending only on a set of mutually commuting quantum observables \cite{GBTr20}.
In addition, upon using the exact factorization $\Upsilon(q,p,x)=\sqrt{D(q,p)}e^{iS(q,p)/\hbar}\psi(x;q,p)$ \cite{AbediEtAl2012,GBTr22}, we observe that the classical phase $S(q,p)$ enters explicitly  the expression \eqref{cden} of the classical density, thereby affecting classical averages. This feature may appear counterintuitive since classical phases should not generally lead to observable effects. Indeed, we do not measure Hamilton-Jacobi functions. One can argue that in the present formalism the single terms in \eqref{cden} do not possess any physical meaning per se, and only the full expression identifies a relevant quantity. Alternatively, one can  modify the current model by  enforcing a gauge principle to treat  classical phases as a gauge freedom; this is the approach recently proposed in \cite{GBTr22,GBTr21}.

In the present context, both the quantum density matrix and the classical distribution may be realized as suitable projections of a hybrid von-Neumann-type operator $\widehat{\cal D}(q,p)$ so that $\rho_c=\operatorname{Tr}\widehat{\cal D}$ and $\hat{\rho}=\int\!\widehat{\cal D}\,\de q \de p$. Explicitly, one has
\[
\widehat{\cal D}=\Upsilon\Upsilon^\dagger-\operatorname{div}(\mathbb{J}\boldsymbol{\cal A}\Upsilon\Upsilon^\dagger)+i\hbar\{\Upsilon,\Upsilon^\dagger\}
\,.
\]
This quantity is needed  to compute quantum-classical expectation values. For example, the total quantum-classical energy is given by $\operatorname{Tr}(\widehat{\cal D}\widehat{H})=\langle\widehat{H}\rangle$ and analogously $\langle\widehat{A}\rangle=\operatorname{Tr}(\widehat{\cal D}\widehat{A})$ for an arbitrary hybrid observable $\widehat{A}(q,p)$.

A typical initial condition for the QCWE is given by a factorized hybrid wavefunction of the type $\Upsilon_0(q,p,x) =\chi(q,p)\Psi(x)$, where $\chi$ and $\Psi$ are quantum and classical wavefunctions, respecitvely. The Koopman wavefunction $\chi$ is chosen in such a way that at the initial time $\rho_c$ is a Gaussian distribution centered at $(q_0,0)$. Then, the expression \eqref{cden} leads to solving a differential equation whose solution leads to \cite{BoGBTr19}
\beq
\Upsilon_0(q,p,x) =  \Psi(x) \,e^{  -i\frac{ pq_{0}}{2\hbar} } \sqrt{\frac{1}{2\pi} \,   \frac{1-(1+\beta H_0)\, e^{-\beta  H_0}}{\beta  H_0^2} },
\qquad\text{ with }\qquad
H_0=\frac12p^2+\frac12(q-q_0)^2
\,.
\label{ICexp}
\eeq
Here, $\beta$ represents an inverse classical temperature. In the fully quantum description, this initial condition corresponds to the tensor product of a quantum state $\Psi(x)$ with a Gaussian wavepacket centered at $(q_0,0)$.
While the function \eqref{ICexp} is suitably normalized,  it possesses a long tail in the phase space which arises from a decay of the type  $1/H_{0}$. The slow decay of the Koopman wavefunction represents a challenging aspect for the numerical implementation, which especially intervenes in the computation of expectation values.

\section{Pure-dephasing systems}\label{sec:puredephasing}

This paper presents a case of study focusing on the interaction of a classical oscillator with a quantum two-level system. In the fully quantum domain, these models were widely studied in the open systems literature \cite{BrPe02}, and are often known in optics as variations of the quantum Rabi \cite{Huang2017,Xie2017} and Jaynes-Cummings models  \cite{Larson2021}.
In the corresponding quantum-classical setting, the hybrid wavefunction is a two-component spinor $\Upsilon=(\Upsilon_+,\Upsilon_-)$.
In particular, we focus on pure-dephasing Hamiltonians of the general type \cite{GuFr18,skinner1,skinner2}
\beq\label{PDHam}
\widehat{H}(q,p)=H_C(q,p)\boldsymbol{1}+(H_I(q,p)+B_0)\widehat{\sigma}_z
\,.
\eeq
Here, $H_C$ is a purely classical Hamiltonian function, while $H_I$  is a quantum-classical coupling term and $B_0$ represents an external magnetic field driving the spin variable. Also,
 $\widehat{\sigma}_z = {\rm diag}(1,-1)$ is the third Pauli matrix and $\boldsymbol{1}$ is the identity matrix.

\subsection{Hybrid wave equation for pure-dephasing Hamiltonians}
For generic nonlinear Hamiltonians in this class, the QCWE  reads
\beq\label{qcwe-generic}
i\hbar\frac{\partial \Upsilon_{\pm}}{\partial t} = i\hbar \{H_\pm,\Upsilon_\pm\} - \mathcal{L}_\pm \Upsilon_\pm
\,,
\eeq
with $H_\pm=H_C\pm(H_I+B_0)$ and $\mathcal{L}_\pm(\mathbf{z}) = \mathbf{z} \cdot \nabla H_\pm(\mathbf{z})/2 - H_\pm(\mathbf{z})$.
Our numerical solution takes advantage of the Madelung representation
$
\Upsilon_\pm = \sqrt{D_\pm}\, \exp{(iS_\pm/\hbar)}
$.
Then, the densities $D_{\pm}$ and phases $S_{\pm}$ obey the following equations in the phase space
\begin{eqnarray}
\partial_t D_{\pm} &=& \{H_\pm,D_\pm\}\,, \label{eq:vlasov1} \\
\partial_t S_{\pm} &=& \{H_\pm,S_\pm\} + \mathcal{L}_\pm. \label{eq:vlasov2}
\end{eqnarray}
These are the transport equations that will be solved numerically to obtain the quantum-classical hybrid results.
The chosen numerical algorithm is a grid-based split-operator technique that advances the functions $D_{\pm}$ and $S_{\pm}$  alternatively in position space $q$ and in momentum space $p$, using a finite volume method \cite{Filbet2001}.
Grid-based computational methods possess good accuracy and stability properties, usually better than methods based on trajectories, although they require a higher computational cost.
More details are given in the Appendix \ref{app:vlasov}. Our initial condition  for  $\Upsilon_{0\pm}=\sqrt{D_{0\pm}}\exp({iS_{0\pm}/\hbar})$
reads
\[
D_{0\pm}=\frac{1}{2\pi}    \frac{1-(1+\beta H_0)\, e^{-\beta  H_0}}{\beta  H_0^2}
\,,\qquad\
S_{0\pm}=-\frac{ pq_{0}}2 \,,
\qquad\text{ with }\qquad
H_0=\frac12p^2+\frac12(q-q_0)^2
\]
which corresponds to \eqref{ICexp} with $\Psi(x)$ replaced by the  spinor $\Psi=(1,1)/\sqrt{2}$.

In this Madelung formulation, the classical density is written as:
\beq
\rho_c(\mathbf{z}) = \sum_\pm \left[ D_\pm + {\rm div} \left( D_\pm\mathbb{J}\nabla S_\pm+\mathbf{z} D_\pm /2\right) \right],
\eeq
with $\int \!\rho_c(\mathbf{z}) d \mathbf{z} = \int [D_{+}(\mathbf{z})  + D_{-}(\mathbf{z})] d \mathbf{z}$, while the quantum density matrix is:
\begin{eqnarray}\nonumber
\hat \rho = \int \Upsilon(\mathbf{z}) \Upsilon^\dag(\mathbf{z}) d\mathbf{z} &=&\!
\int d\mathbf{z}  \left(
 \begin{matrix}
|\Upsilon_{+}|^2 & \Upsilon_{+} \Upsilon_{-}^* \\
\Upsilon_{+}^* \Upsilon_{-} & |\Upsilon_{-}|^2
 \end{matrix} \right)  \\
 &=&\! \int d\mathbf{z}  \left(
 \begin{matrix}
D_{+} & \sqrt{D_{+} D_{-}} e^{i\Delta S/\hbar} \\
\sqrt{D_{+} D_{-}} e^{-i\Delta S/\hbar}  & D_{-} \,,
 \end{matrix}
 \right)
\end{eqnarray}
where $\Delta S = S_{+}-S_{-}$ is the phase difference.

\subsection{Inadequacy of the Ehrenfest model and other features\label{sec:features}}
Despite their intrinsic simplicity, pure-dephasing systems represent a relevant benchmark test case for two main reasons.

First, pure-dephasing systems cannot be adequately modeled by the Ehrenfest equations \eqref{Ehrenfest}, since in that case the classical motion decouples entirely from the quantum dynamics. This is easy to see by noticing that the initial condition $\langle\psi|\widehat{\sigma}_z\psi\rangle=0$ is preserved in time, so that the classical evolution in \eqref{Ehrenfest} simply becomes
\[
\partial_t D+\{D,H_C\}=0
\,.
\]
In particular, this indicates the complete absence of quantum backreaction on the classical motion, so that, when $H_C=(p^2+q^2)/2$ the classical system undergoes trivial rotations in  phase space, in obvious contrast with the behavior arising from the fully quantum treatment.

Another relevant property of pure-dephasing systems is that, whenever $H_C$ and $H_I$ are quadratic functions, the QCWE can be solved exactly \cite{BoGBTr19,GBTr20} thereby providing important information to test the numerical results. In this particular case the evolution of the classical density \eqref{cden} is shown  to be identical to the Wigner oscillator dynamics predicted by the corresponding fully quantum problem. Indeed, since the two components  $\Upsilon_\pm$ of the hybrid wavefunction decouple entirely, one shows that the two densities $$\rho_\pm=|\Upsilon_\pm|^2+\frac12\operatorname{div}(\mathbf{z}|\Upsilon_\pm|^2)+\hbar\operatorname{Im}\{\Upsilon^*_\pm,\Upsilon_\pm\}$$
obey a classical Liouville equation:
$\partial_t \rho_\pm=\{H_\pm,\rho_\pm\}$. 
Notice that  the densities $\rho_\pm$ are transported  along the characteristics of the time-independent vector field $(\partial_pH_\pm,-\partial_qH_\pm)$ and thus they remain strictly positive at all times. Then, the sign of the classical density $\rho_c=\rho_++\rho_-$ is also strictly preserved, thereby indicating the absence of wavefunction nodes that may instead appear in the general case of quantum dynamics.

For the corresponding quantum problem, the Wigner function of the oscillator wavefunction satisfying
\beq\label{quantumproblem}
i\hbar\partial_t\psi=\widehat{H}_C\psi+(\widehat{H}_I+B_0)\widehat{\sigma}_z\psi
\eeq
is given by $W=W_++W_-$. Here, $W_\pm$ is the Wigner transform of $\psi_\pm$ and obeys $\partial_t W_\pm=\{\!\{H_\pm,W_\pm\}\!\}$, where $\{\!\{\cdot,\cdot\}\!\}$ is the Moyal bracket. Then, if $H_\pm$ is a quadratic function the Moyal bracket reduces to the Poisson bracket, thereby recovering exactly the same classical flow that arises from the QCWE. Notice how this result is in direct contrast with that produced by the Ehrenfest model, for which the oscillator dynamics decouples completely and becomes harmonic.

In the following sections, we will compare the results of the QCWE with those arising from the fully quantum treatment of the Hamiltonian \eqref{PDHam}. For more details on the quantum dynamics, see Appendix \ref{app:schrodinger}. In particular, we will consider different cases of $H_C$ and $H_I$ corresponding to harmonic oscillator motion with linear and quadratic coupling (Sec. \ref{sec:harmo}) and two types of anharmonic potentials and coupling (Sec. \ref{sec:anharmo}). Regimes with both constant and time-dependent parameters will be investigated.

\section{Harmonic motion}\label{sec:harmo}
This section presents the results obtained from the numerical implementation of the QCWE for  pure-dephasing Hamiltonians of the type \eqref{PDHam}, in the special case when
\beq\label{oscham}
H_C=\frac12\left(p^2+\omega^2q^2\right)
\eeq
 and $H_I$ is either linear or quadratic. This hybrid Hamiltonian represents the pure-dephasing interaction of a harmonic oscillator that is coupled linearly or quadratically with a two-level system. In particular, we will compare the results obtained with the QCWE to those obtained from the fully quantum evolution.
 In the last part of this section, we will also study the case in which either the oscillator frequency $\omega$ or the parameters in $H_I$ depend explicitly on time.

 Notice that, as shown before, in the case of Hamiltonians that are at most quadratic,  the classical density evolution obtained from the QCWE coincides exactly with the Wigner distribution dynamics that is obtained from the fully quantum treatment. Thus, the present study will only consider the dynamics of the Bloch vector describing the quantum two-level subsystem.

\subsection{Linear coupling}
Here, we consider in \eqref{PDHam} a linear interaction Hamiltonian of the type
\beq\label{lincoupham}
H_I=\gamma q
\eeq
and the parameter values:
\[
\omega=1,\qquad
\gamma=0.5,\qquad
B_0=0.2.
\]
The initial condition is given by \eqref{ICexp} with initial shift $q_0=0$ and inverse temperature $\beta=2$.
Units in which $\omega=\hbar=m = 1$ are used throughout this paper. A fully quantum system corresponding to  \eqref{PDHam}, \eqref{oscham}, and \eqref{lincoupham} was discussed in \cite{ReSiSu96} in the context of electron-phonon coupling. In the case $B_0=0$, the corresponding quantum Hamiltonian  identifies the simplest version of the Jahn-Teller problem in chemical physics, known as $E\otimes\beta$ Jahn-Teller model \cite{OBCh93}.

\begin{figure}[h]
\center
\includegraphics[scale=.25]{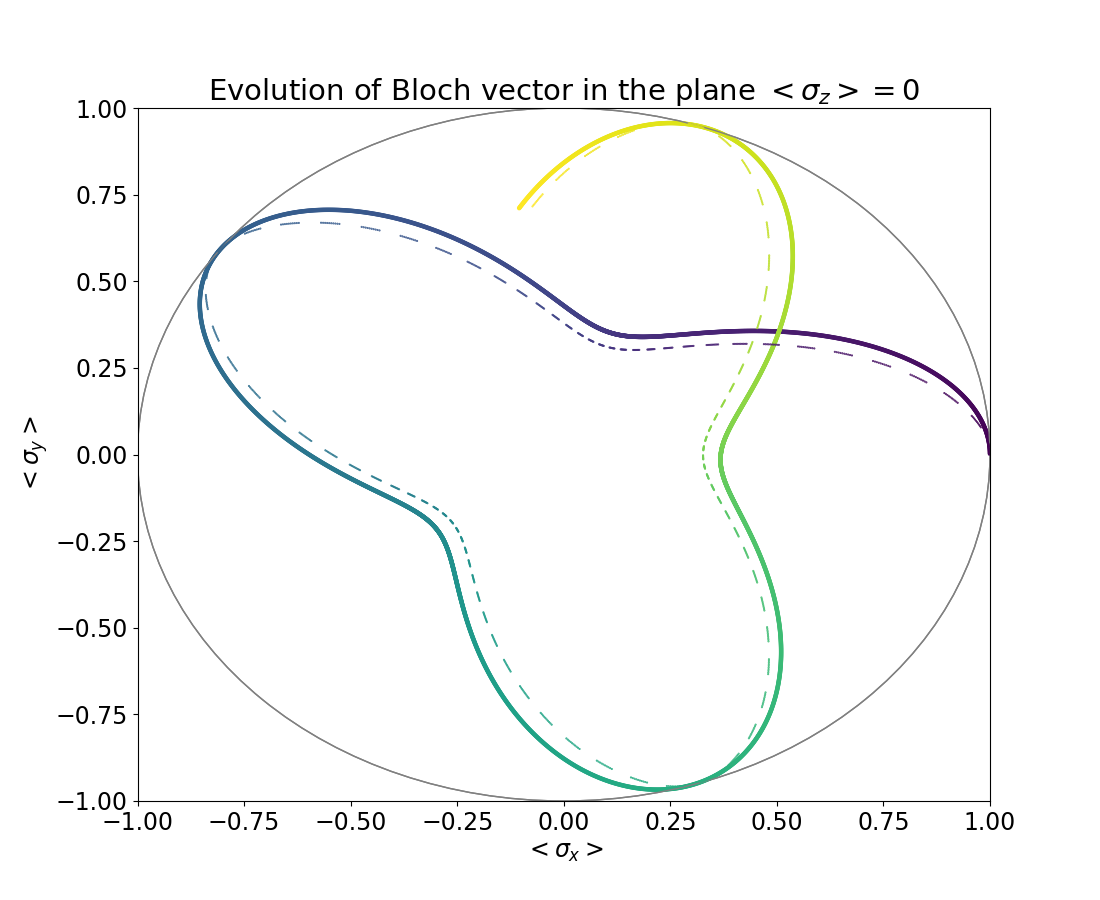}
\includegraphics[scale=.25]{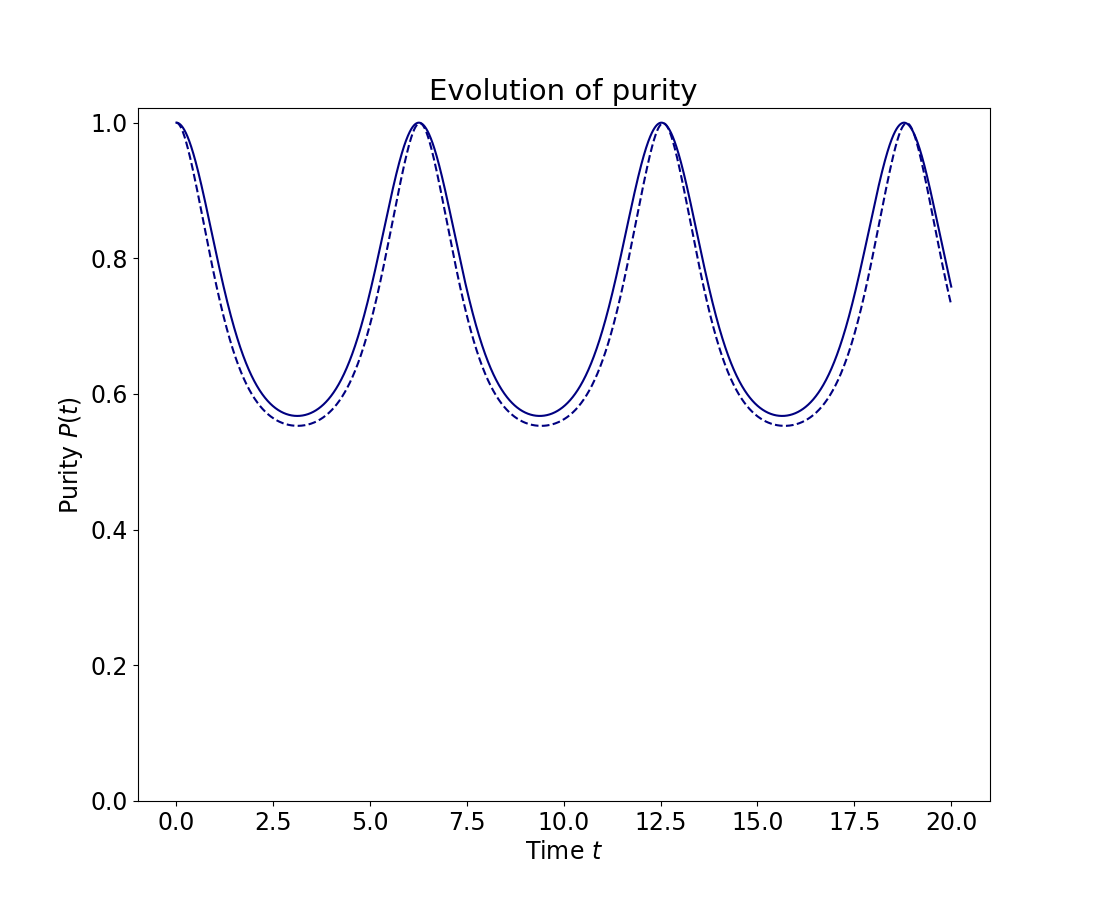}
\caption{{\it Dynamics of the quantum subsystem (linear coupling)}. The left panel shows the evolution of the Bloch vector in the equatorial disk $\langle \sigma_z \rangle =0 $: the purple color corresponds to earlier times, while yellow indicates later times $t=20$. The dashed line corresponds to the hybrid evolution, while the thick solid line represents the fully quantum dynamics. The right panel shows the evolution of purity with the same convention. The same graphical conventions and color code will be used throughout the remaining figures.}
\label{SpinLin}
\end{figure}

Figure \ref{SpinLin} shows the Bloch vector dynamics along with purity evolution, where purity is given by $P(t)=\operatorname{Tr}\hat\rho^2$.
Since in the present pure-dephasing case $\langle \sigma_z \rangle $ is conserved, the motion occurs in the plane $\langle \sigma_z \rangle =0 $.
While the former gives an overview of the quantum subsystem dynamics (spin dynamics), the latter quantifies decoherence. Notice the good agreement between the hybrid and fully quantum results, although we see the slightly lower purity level attained by the hybrid system. The latter feature means that the decoherence induced on the two-level subsystem is higher in the case of an interaction with a classical oscillator, compared to a fully quantum oscillator.

\begin{figure}
\center
\includegraphics[scale=.35]{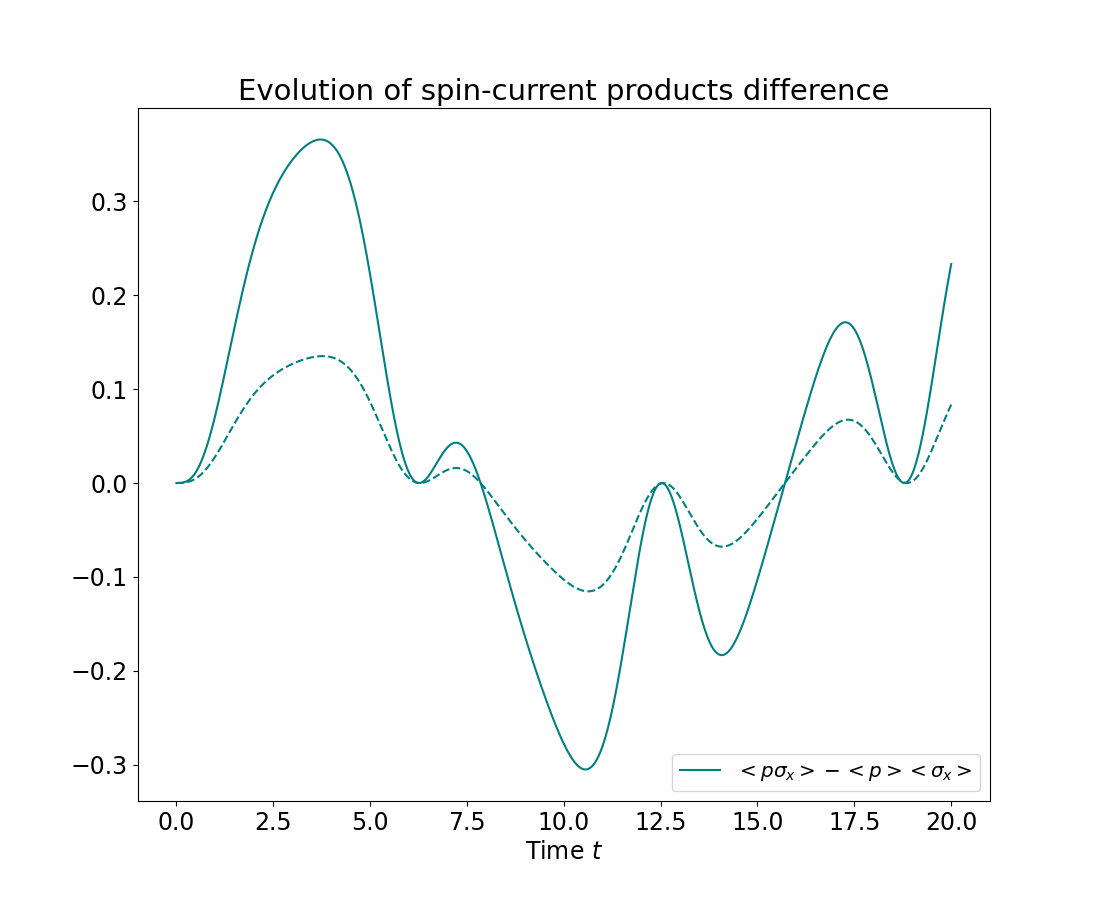}
\caption{{\it Spin-momentum correlations (linear coupling)}. The pale blue curves indicate the evolution of the spin current correlation $\langle p\widehat\sigma_x\rangle - \langle p\rangle \langle\widehat\sigma_x\rangle$, for the quantum evolution (solid curve) and the hybrid evolution (dashed curve). Note that in this case $\langle p\rangle=0$, because no initial shift was present ($q_0=0$).
}
\label{SpinCurrentLin}
\end{figure}

Another possibly relevant quantity is the spin current $\langle p\widehat\sigma_x\rangle$. Here, we are particularly interested in the quantum-classical spin-momentum correlations $\langle p\widehat\sigma_x\rangle-\langle p\rangle\langle\widehat\sigma_x\rangle$ whose time evolution is represented in Figure \ref{SpinCurrentLin}.
The spin-momentum correlations achieve lower absolute values in the hybrid case.
In other words, the  correlations generated in time by the fully quantum dynamics are higher than the quantum-classical correlations achieved during the hybrid evolution. In addition, we  notice that when the quantum state turns back pure, the spin-momentum correlations vanish for both quantum and hybrid dynamics.

Overall, we found that the Bloch dynamics resulting from the QCWE is in good agreement with the predictions of the fully quantum theory, while the correlations between the oscillator and the two-level system reach higher values in the fully quantum case. We recall that the harmonic oscillator evolution is exactly the same in the two cases.

\subsection{Quadratic coupling\label{sec:quadcoup}}
This section focuses on pure-dephasing systems of the type \eqref{PDHam}, where $H_C$ is the harmonic oscillator Hamiltonian \eqref{oscham} and the interaction term is given by
\beq\label{quadcoupham}
H_I=\frac\eta2(q^2-p^2) ,
\eeq
with parameters
\[
\omega=1,\qquad
\eta=0.5,\qquad
B_0=0.3.
\]
In the absence of momentum coupling, the corresponding quantum case was considered in \cite{ReSiSu96}.
In the absence of both momentum coupling and magnetic field, an analogue quantum-classical system was approached by the QCWE in \cite{BoGBTr19}. In the case considered here, the quantum correspondent of the coupling Hamiltonian \eqref{quadcoupham} has made its appearance in the context of the two-photon Rabi dynamics \cite{AlScBr11}. The pure-dephasing limit was considered in \cite{EmBi02,Ng99,Trav12}, although in those cases the magnetic field $B_0$ appearing in \eqref{PDHam} is absent.
While the inverse temperature  of the initial condition is fixed ($\beta=2$), here we will study both cases $q_0=0$ and $q_0=1$ to observe the effects of an initial shift in position.

It is perhaps interesting to mention that, if $q_0=0$, switching off the external magnetic field $B_0$ leads to an absence of Bloch rotation in both the quantum and the hybrid cases. This can be seen explicitly for the QCWE equation upon using the polar form $\Upsilon_\pm=\sqrt{D_\pm}e^{iS_\pm/\hbar}$  in the corresponding wave equations. Then, the QCWE leads to the expectation value expressions
\begin{align*}
\big(
\langle\widehat\sigma_{x}\rangle
,
\langle\widehat\sigma_{y}\rangle
\big)&=2\int\!\sqrt{D_+ D_-}\big(\cos(\Delta S/\hbar),-\sin(\Delta S/\hbar)\big)\,\de q\de p
\,,
\end{align*}
where $\Delta S=S_+-S_-$. The latter quantity vanishes at all times due to the initial condition \eqref{ICexp} and the particular form $\partial_t \Upsilon-\{ H_C\boldsymbol{1}+H_I\widehat{\sigma}_z, \Upsilon\}=0$ of the QCWE when $B_0=q_0=0$. Thus, $\langle\widehat\sigma_y\rangle=0$ remains true at all times and the Bloch vector simply oscillates in amplitude along the $x-$direction. The same peculiar behavior is also observed in the simulations of the fully quantum dynamics, although its analytical justification is less obvious.

\begin{figure}
\center
\includegraphics[scale=.25]{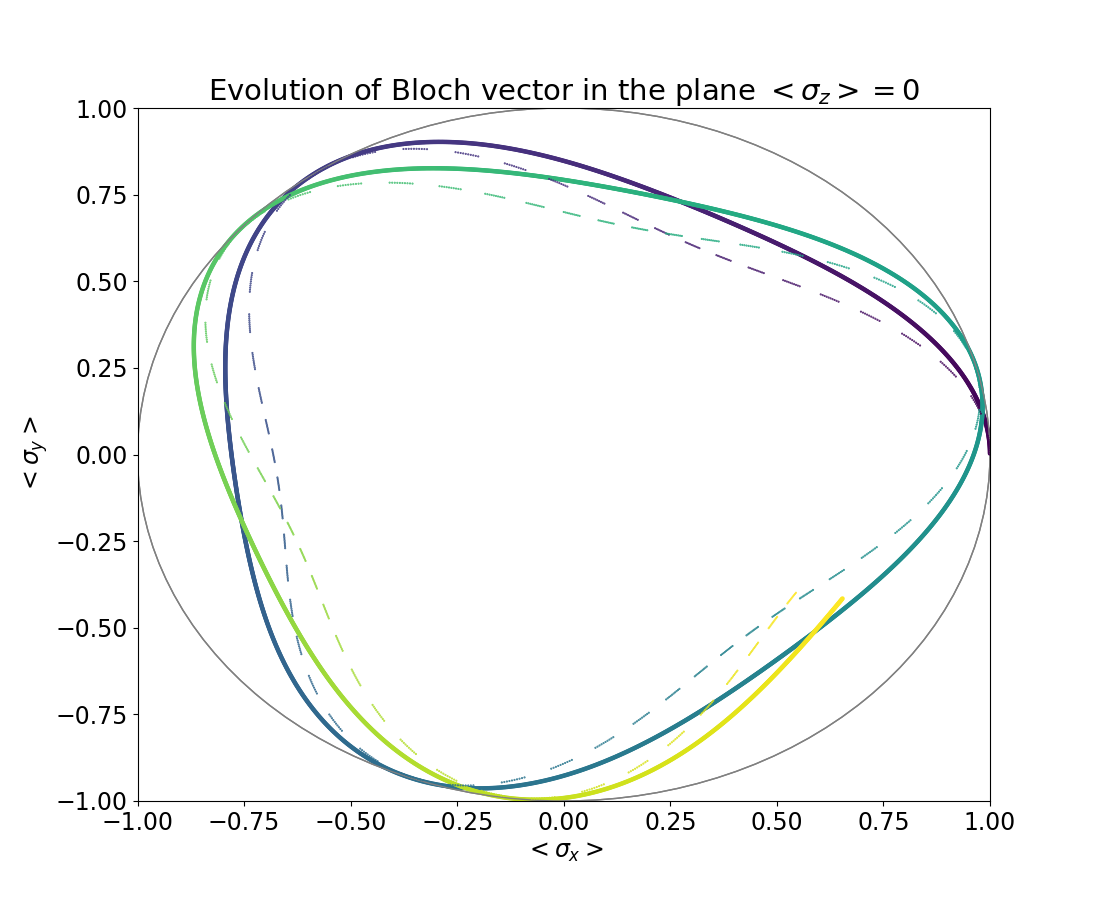}
\includegraphics[scale=.25]{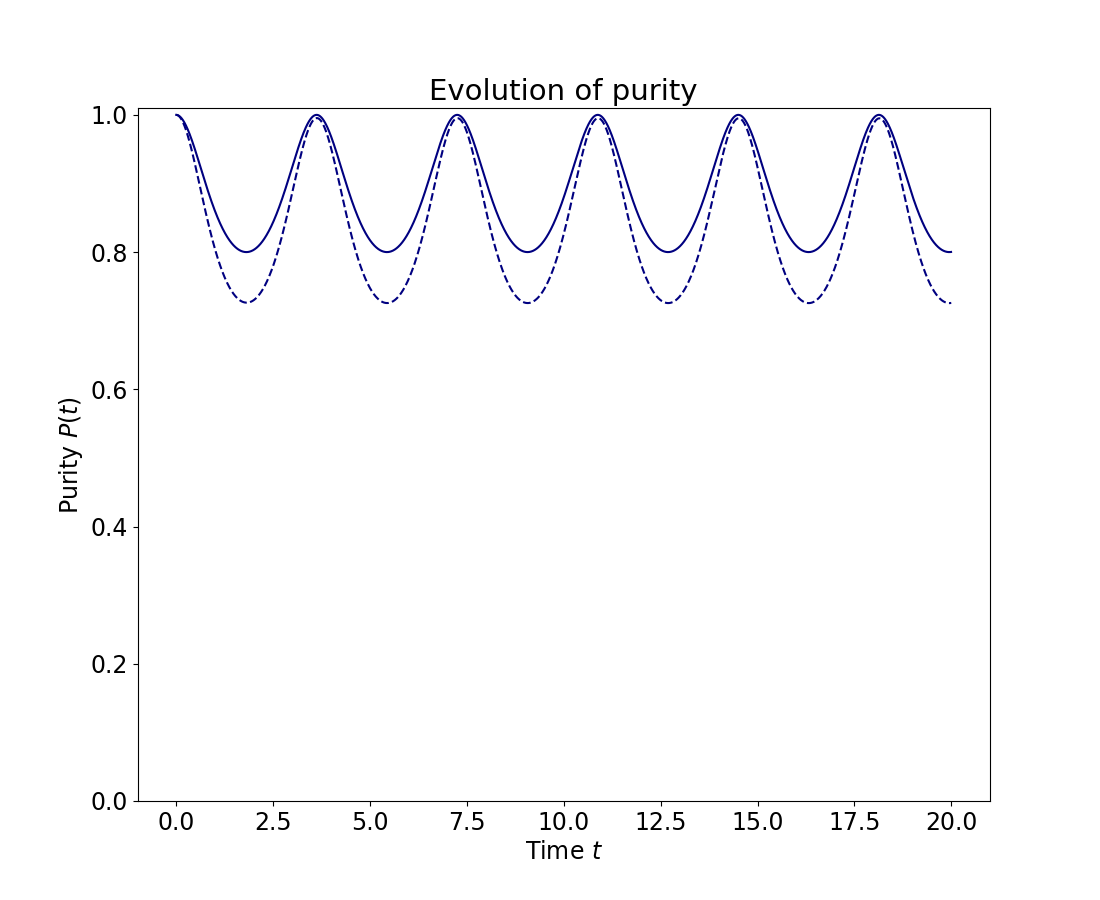}\\
\includegraphics[scale=.25]{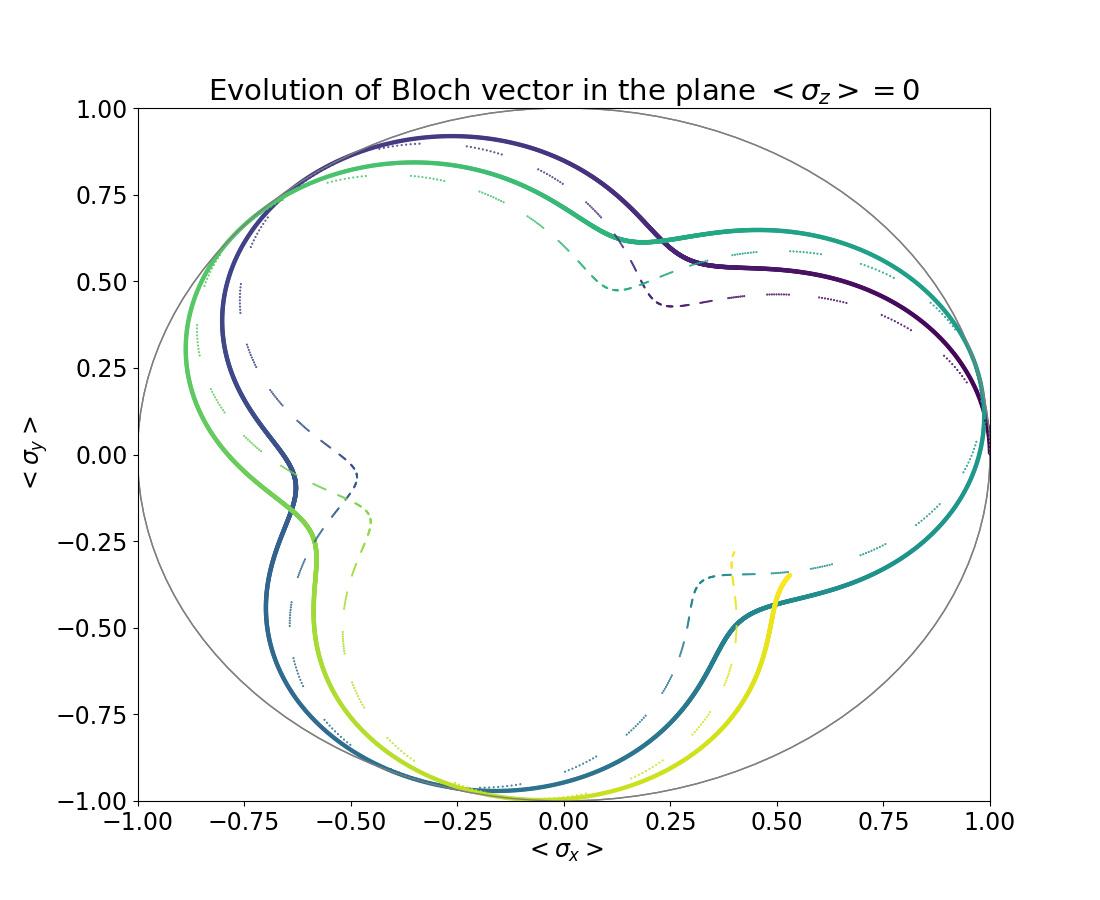}
\includegraphics[scale=.25]{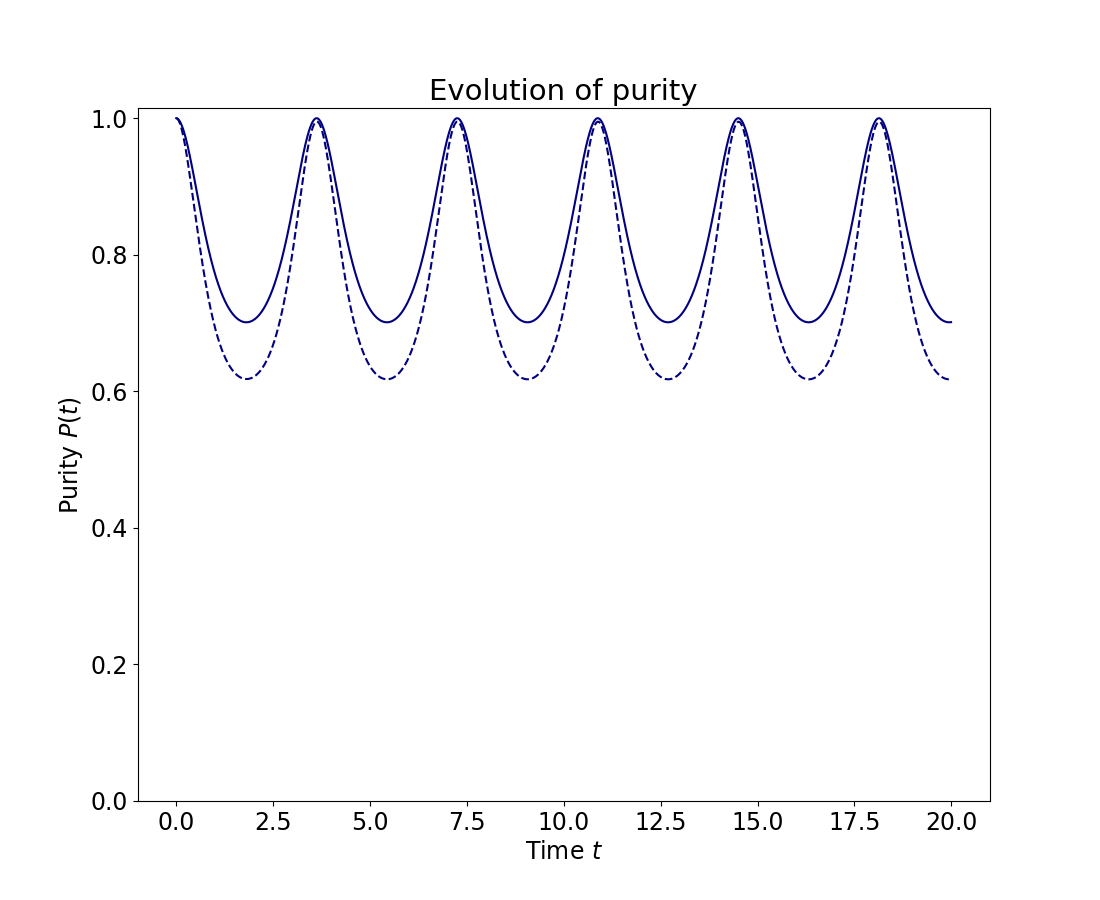}
\caption{{\it Dynamics of the quantum subsystem (quadratic coupling)}. Bloch and purity dynamics corresponding to the quadratic coupling in \eqref{quadcoupham}.  The upper panels refer to the case $q_0=0$, while the lower panels represent the case $q_0=1$.}
\label{SpinQuad}
\end{figure}

The remainder of this section deals with the case of a nonzero magnetic field, $B_0=0.3$.
Figure \ref{SpinQuad} illustrates the dynamics of the two-level subsystem. Independently of the presence of an initial shift, we observe essentially the same features that had already appeared in the case of linear coupling, although now the difference in purity reaches slightly higher values, meaning that the quadratic coupling leads to stronger  decoherence effects in the hybrid case.

\begin{figure}
\center
\includegraphics[scale=.35]{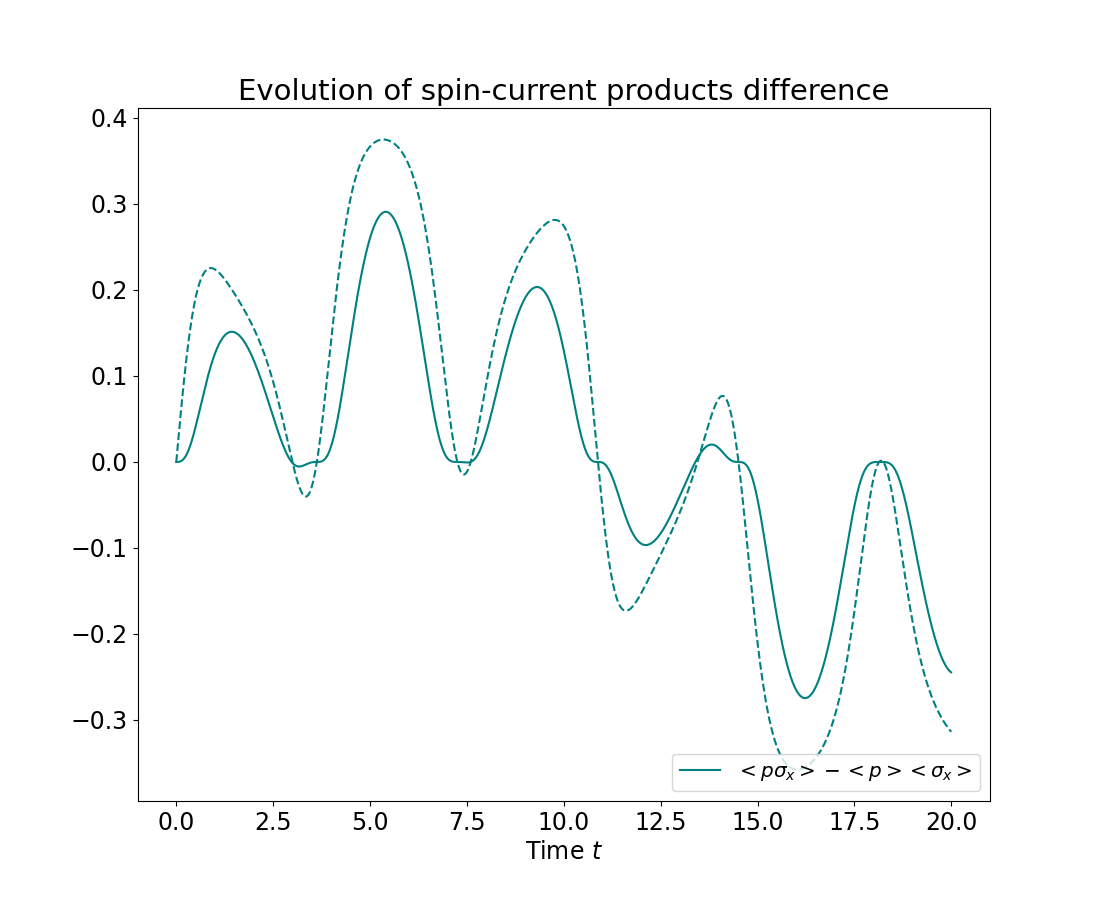}
\caption{{\it Spin-momentum correlations  (quadratic coupling)}.
Same as Figure \ref{SpinCurrentLin}, although now the curves are obtained in the case of quadratic coupling with an initial position displacement $q_0=1$. Notice that $\langle p\rangle$ does not vanish in this case.
}
\label{SpinCurrentQuad}
\end{figure}

In the presence of an initial shift $q_0=1$, the spin current may again be adopted to illustrate the correlation dynamics between the oscillator and the two-level system (see Figure \ref{SpinCurrentQuad}). In contrast with the case of linear coupling (Figure \ref{SpinCurrentLin}), the hybrid correlations are now higher than those observed in the full quantum dynamics. We recall, however, that the linear coupling case in Figure \ref{SpinCurrentLin} did not involve any initial shift, so the two results are not directly comparable.
Again, we  notice that, when purity reaches its maximum,  spin-momentum correlations vanish and this happens  for  both quantum and hybrid dynamics.

In summary, the case of quadratic coupling does not lead to any substantial differences with respect to the conclusions found when the coupling is linear. The main difference resides in a slightly bigger difference in purity, whose quantum predictions are however still qualitatively reproduced.

\subsection{Time-dependent parameters}

In this section, we keep the form of the Hamiltonians \eqref{oscham} and \eqref{quadcoupham} and consider the separate cases where either the oscillator frequency $\omega$ or the coupling constant $\eta$ are dependent on time.  As we will see, while the case of a variable coupling yields again a good qualitative agreement, the case of variable frequency leads to a very peculiar quantum dynamics that escapes from the QCWE description, which instead reproduces the quantum oscillator dynamics exactly. Notice that a variable oscillator frequency is a less realistic situation than the case of a variable coupling strength.

We start by considering the following temporal expression for $\omega(t)$:
\beq
\omega(t) = 1  + {1 \over 2} \Delta\omega\left[\tanh{(t-t_1)} - \tanh{(t-t_2)}\right]
,
\label{tdfreq}
\eeq
with $\Delta \omega = 0.4$, $t_1=20$, and $t_2=60$.
In this case, we ignore the effects of an initial displacement $q_0$ and we consider a much lower magnetic field $B_0=0.02$ in order to emphasize the effects of the time-dependence. The inverse temperature is still $\beta=2$  and the coupling strength is $\eta=0.5$.
\begin{figure}[h]
\center
\includegraphics[scale=.27]{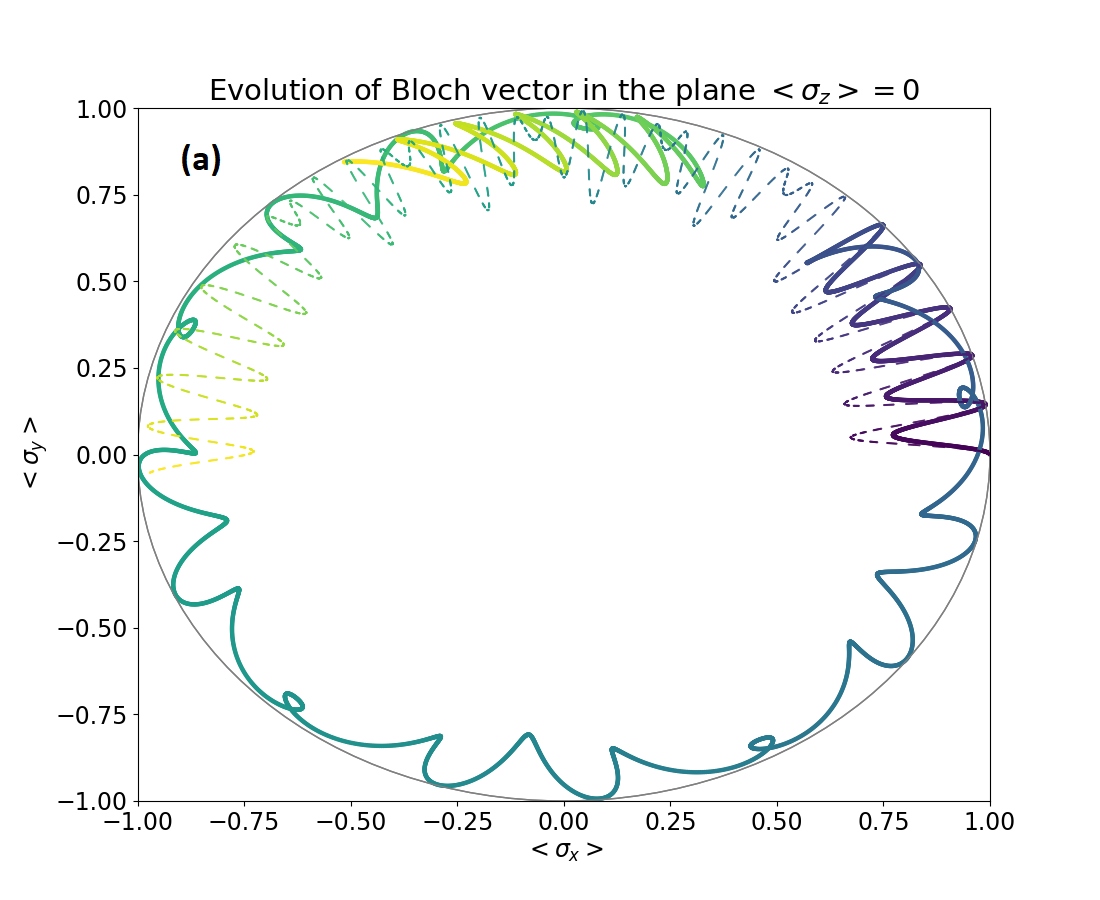}
\includegraphics[scale=.27]{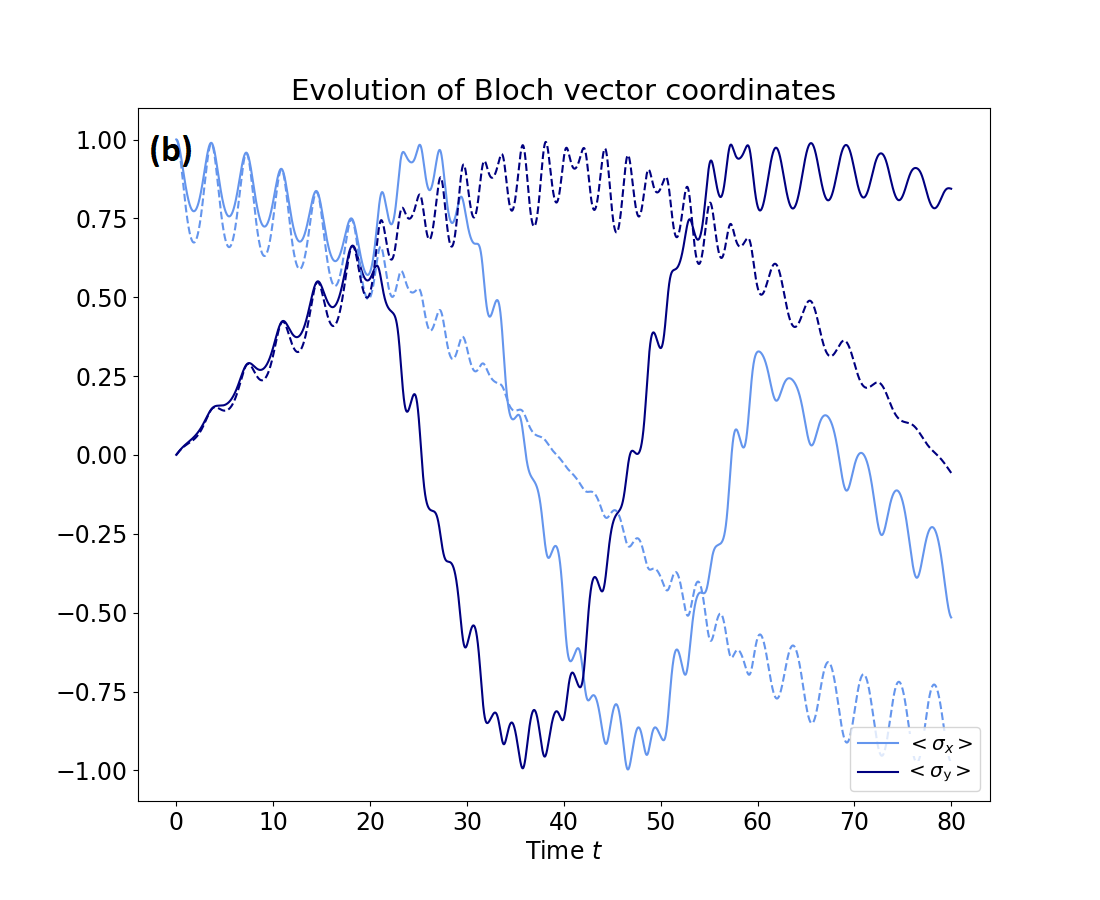}
\includegraphics[scale=.27]{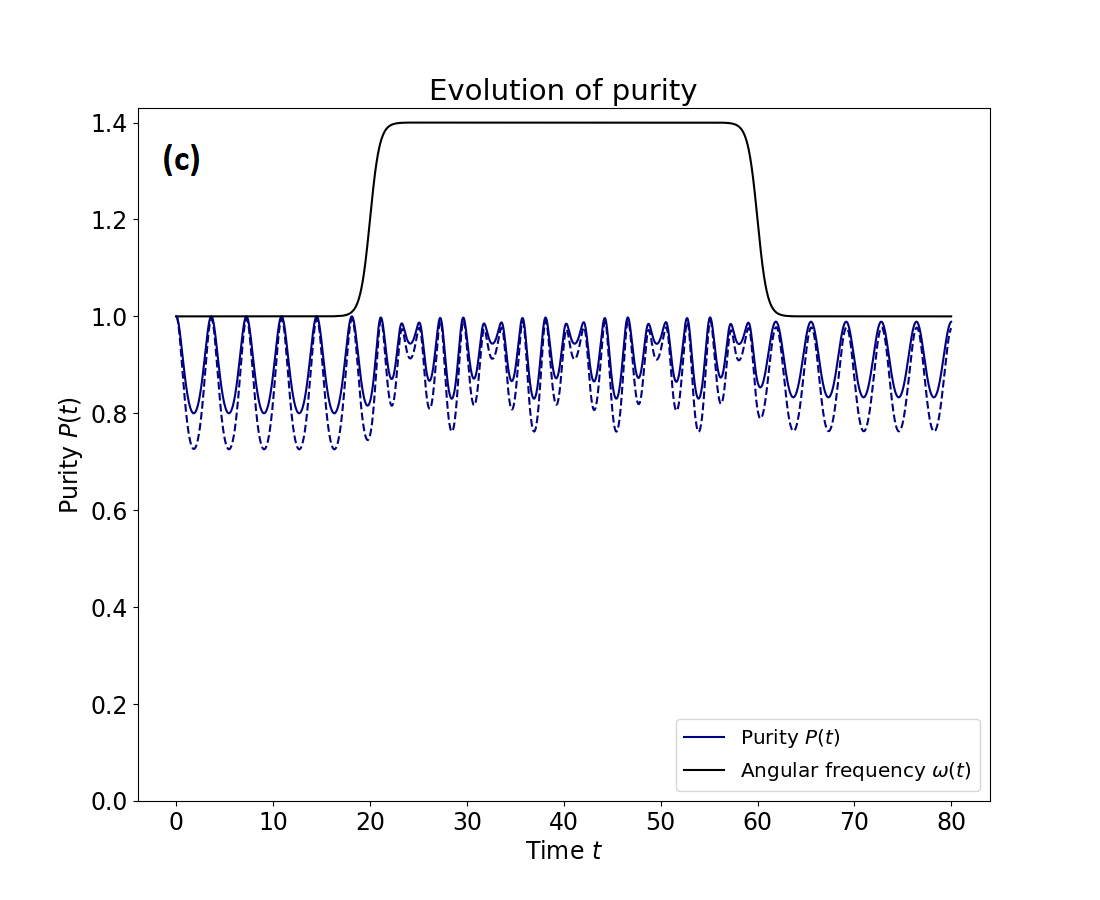}%
\caption{{\it Dynamics of the quantum subsystem}.  (a) Bloch dynamics in the equatorial disk $\langle \sigma_z \rangle =0 $; (b) dynamics of the single components $\langle \sigma_x \rangle$ and  $\langle \sigma_y \rangle$; and (c) purity dynamics
corresponding to the Hamiltonians \eqref{oscham} and \eqref{quadcoupham}. The time-dependent frequency $\omega(t)$ in \eqref{tdfreq} is presented in panel (c) as a black solid line.
}
\label{SpinQuadPulsVar}
\end{figure}
This case exhibits a very different qualitative behavior of the Bloch rotation between the hybrid and fully quantum cases,  while purity is still satisfactorily reproduced by the QCWE. In particular, the Bloch rotation in the hybrid case is entirely determined by the magnetic field, while the fully quantum case exhibits an atypical Bloch dynamics that is triggered by a nonzero relative phase  of  the spinor components. As shown in Figure \ref{SpinQuadPulsVar}, at the time when the frequency develops a   derivative  ($t_1 = 20$), the positive (counterclockwise) Bloch rotation  observed in the fully quantum case reverses its sign and becomes very fast  until the frequency returns to its original value. After that point in time  ($t_2 = 60$), the Bloch rotation slows down and turns back positive. On the other hand, this remarkable effect  cannot be captured by the hybrid QCWE dynamics, for which the rotation remains  slow and positive at all times. Instead, the classical density $D=\|\Upsilon\|^2$ resulting from the QCWE reproduces the quantum  dynamics of the time-dependent oscillator exactly, as discussed in Section \ref{sec:features}

\begin{figure}[h]
\center
\includegraphics[scale=.3]{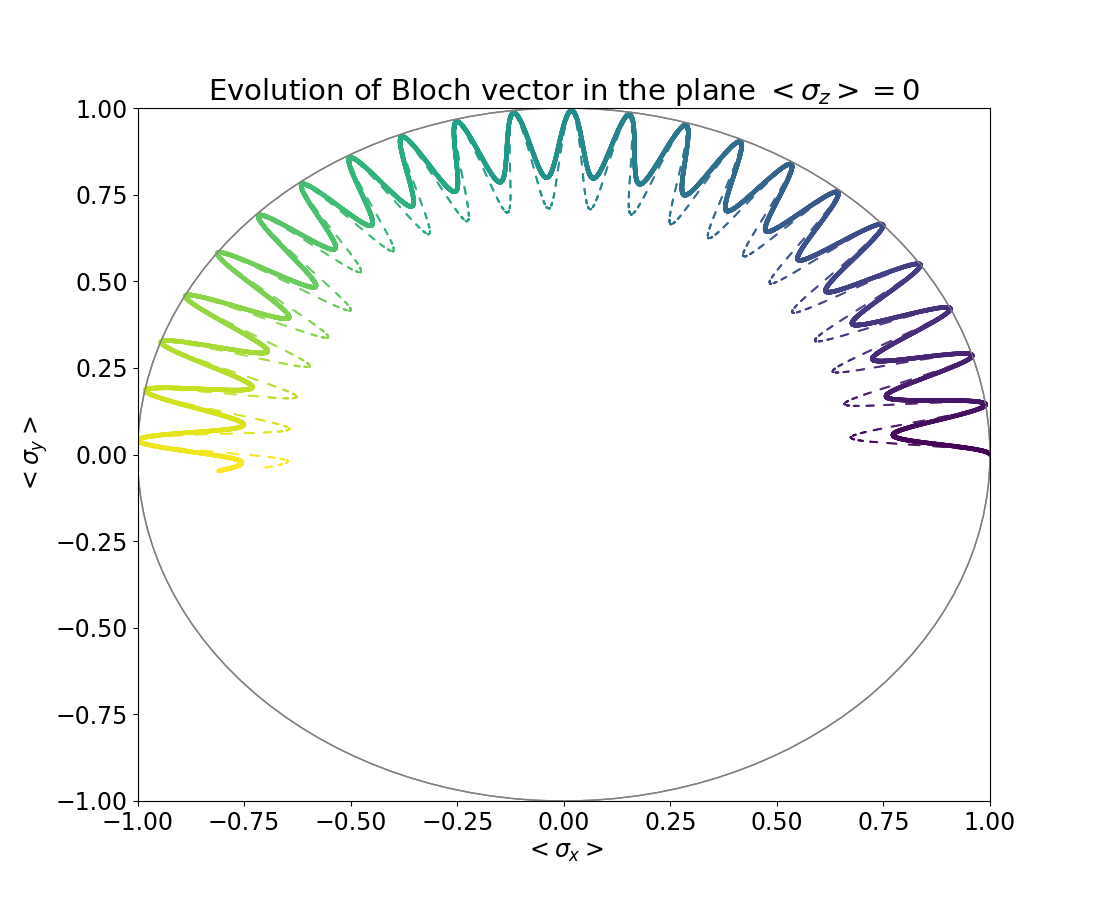}
\includegraphics[scale=.3]{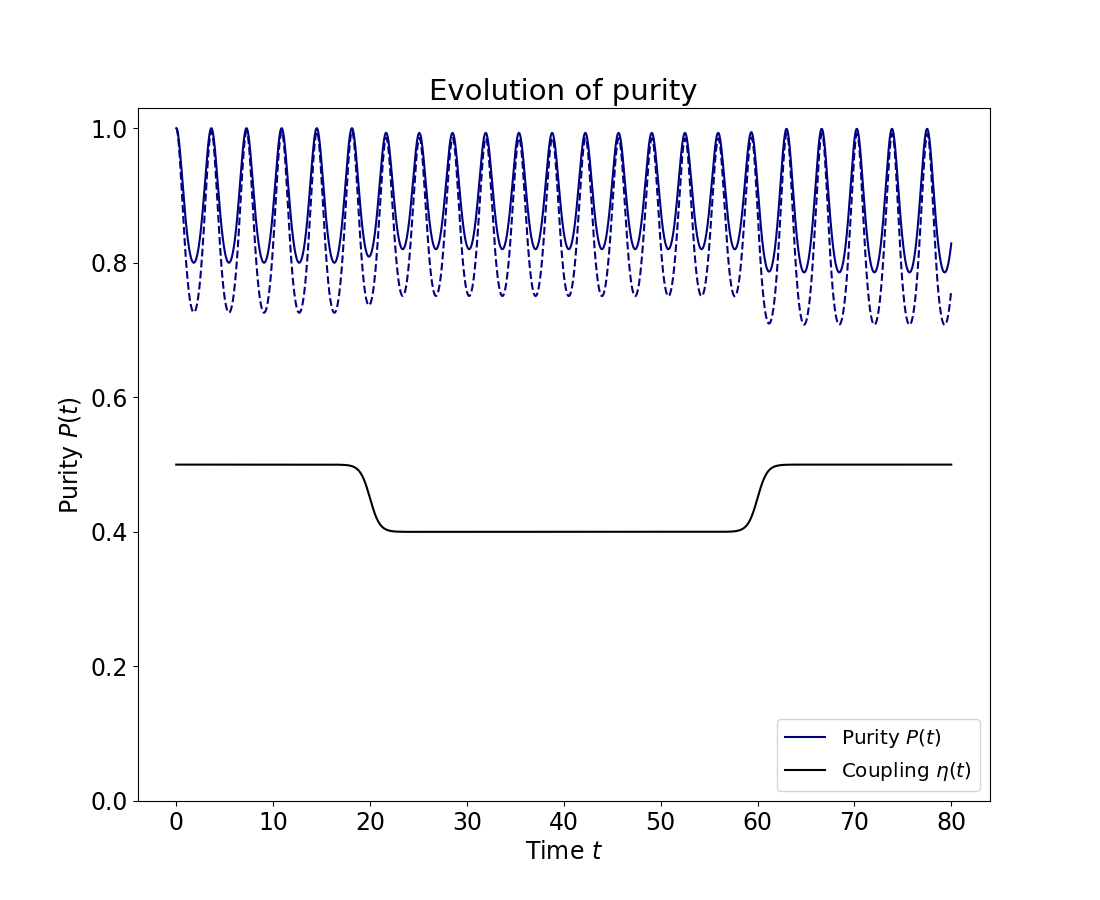}
\caption{{\it Dynamics of the quantum subsystem}.  Bloch (left panel) and purity (right panel) dynamics corresponding to the Hamiltonians \eqref{oscham} and \eqref{quadcoupham}, with a time-dependent coupling constant $\eta(t)$  whose  evolution is presented in the right panel as a black solid line (see main text for the analytical formula). The results from the hybrid (dashed lines) and the fully quantum (solid lines) dynamics exhibit overall qualitative agreement.}
\label{SpinQuadCoupVar}
\end{figure}

In the second simulation, we take a constant oscillator frequency  ($\omega = 1$) and a time-dependent coupling coefficient $\eta(t)$ given by
$$\eta(t) = \eta_0 + {1 \over 2}\Delta\eta\left[\tanh{(t-t_1)} - \tanh{(t-t_2)}\right],$$
with $\eta_0 = 0.5$ and $\Delta \eta = -0.1$.
The results are illustrated in Figure \ref{SpinQuadCoupVar}.
Contrarily to the previous situation (constant coupling and varying frequency), here the QCWE reproduces rather well both the Bloch rotation and the purity evolution. Once again, as in earlier cases, the purity predicted by the hybrid dynamics reaches slightly lower values compared to the fully quantum case.

\section{Anharmonic motion}\label{sec:anharmo}
For a Hamiltonian that is at most quadratic, the oscillator dynamics is identical in the quantum and hybrid cases, as was shown in Sec. \ref{sec:puredephasing}. In such cases,  we observed a good agreement  also for the quantum two-level subsystem, although distinctive differences emerged in the case of a variable oscillator frequency.

Here, we extend our study to the more challenging scenario of anharmonic motion, for which the quantum and classical dynamics differ at long times. We recall that, as discussed in Section \ref{sec:features}, the Ehrenfest model \eqref{Ehrenfest} is inadequate also for anharmonic pure-dephasing dynamics since it leads to a completely decoupled classical evolution. This problem is absent in the QCWE and here we  present results for two cases, where  either the classical Hamiltonian $H_C(q,p)$ or the coupling term $H_I(q,p)$ involve some nonlinear terms. Two types of nonlinear correction terms are considered: a  quartic term  and a rational term of the type $\alpha q^4/(1+\alpha q^2)$. The latter belongs to a wider class of potentials known as \emph{Mitra potentials}  \cite{carinena,Mitra}. Among others, Mitra  potentials were  used in the calculation of bond state spectra; see \cite{RoJaPr08} for this and other occurrences of Mitra potentials.

\subsection{Quartic potential correction}
In this section, we modify the hybrid dynamics in Section \ref{sec:quadcoup} by adding a quartic potential correction, that is
\beq
 H_C(p,q) = \frac{p^2+ \omega^2 q^2}{2}+ \epsilon\,\frac{q^4}{4},
 \label{quarticham}
\eeq
while the coupling terms remain quadratic:
$ H_I(p,q) =  {1 \over 2}\eta (q^2-p^2) +B_0$. We recall that the quantum dynamics arising from a quartic potential term differs substantially from its classical counterpart \cite{Manfredi1996}. Here, we use the following parameters: oscillator frequency $\omega=1$, quadratic coupling $\eta=0.3$, external field $B_0=0.1$, and correction parameter $\epsilon=0.01$. The initial condition is still the one of Eq. \eqref{ICexp}, with inverse temperature $\beta=2$.

\begin{figure}
\center
\includegraphics[scale=.3]{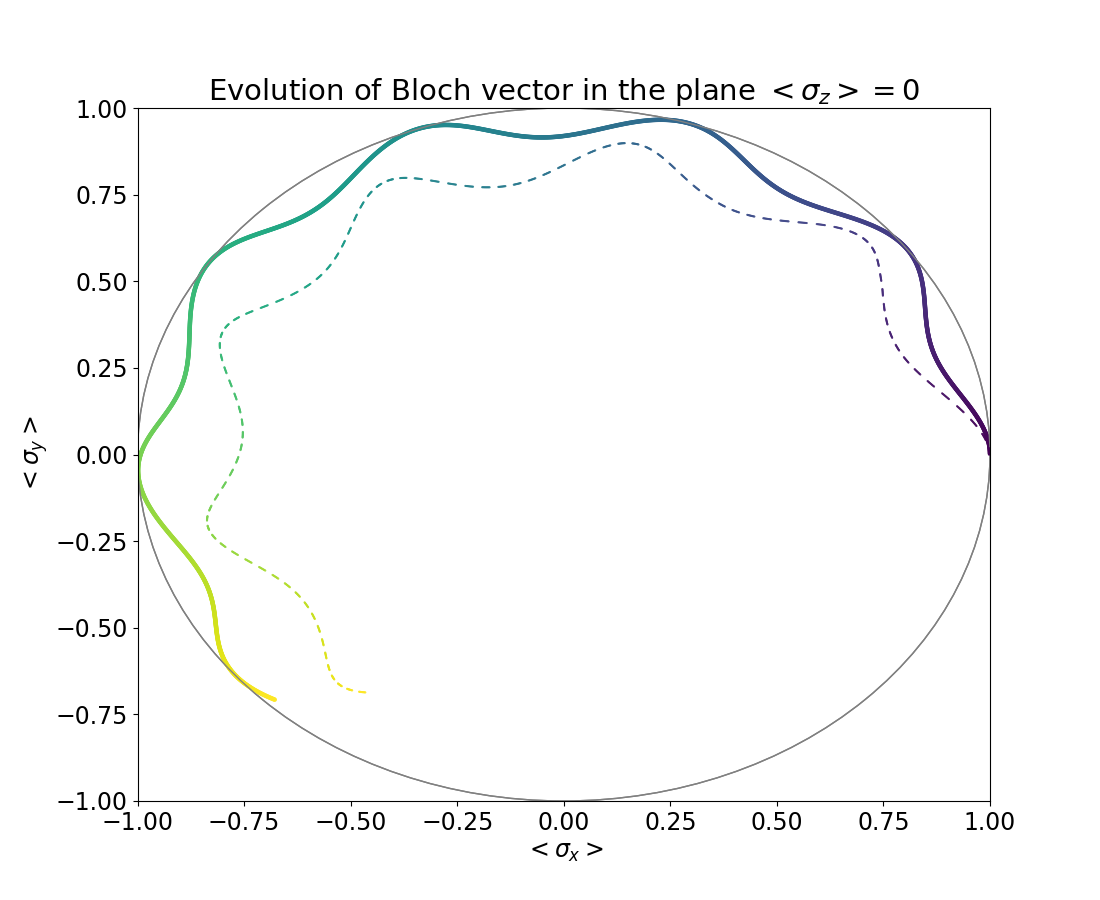}
\includegraphics[scale=.3]{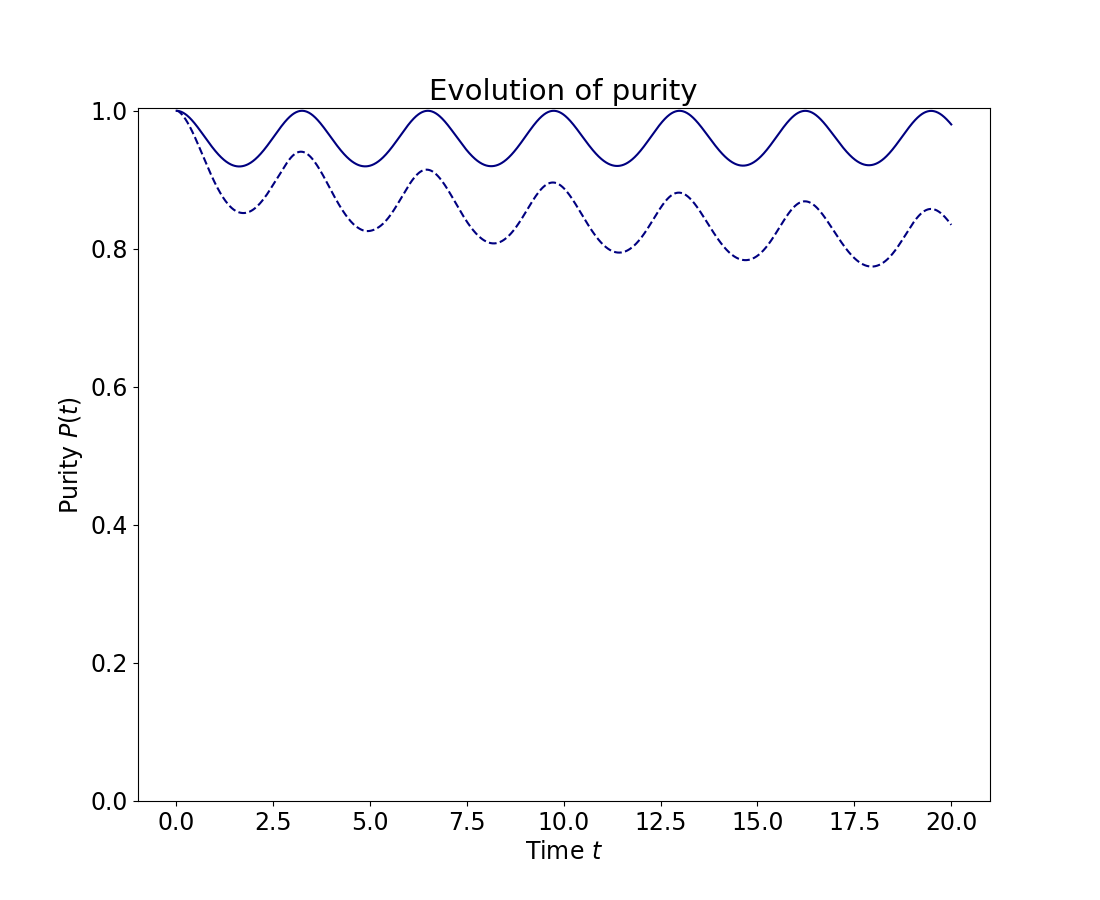}
\caption{{\it Dynamics of the quantum subsystem (quartic potential)}. Bloch and purity dynamics (left and right panel, respectively) corresponding to the pure-dephasing dynamics for  a harmonic oscillator  with a quartic potential correction.
}
\label{QuarticPot}
\end{figure}

The results for the Bloch  and  purity dynamics of the quantum subsystem are displayed in Fig. \ref{QuarticPot}. The purity oscillates while decreasing to a lower value in the hybrid case with respect to the fully quantum case. This result should be compared to the one obtained for a purely quadratic Hamiltonian, see Fig. \ref{SpinQuad}: in that case, the hybrid purity was following more closely the fully quantum one, both returning to values very close to unity. In other words, in this case of quartic potential correction we observe that the decoherence in  the hybrid dynamics tends to become more and more pronounced over time, while its overall average levels do  not seem to vary in the fully quantum case.
Also, we notice a faster Bloch rotation in the hybrid case with respect to the fully quantum result.

Due to the slow decay of the hybrid wavefunction \eqref{ICexp} and the rapid growth of the quartic potential $\sim q^4$, the implementation of the hybrid dynamics corresponding to the classical Hamiltonian \eqref{quarticham}  is particularly challenging and the long-time simulations do not allow a satisfactory energy conservation. In order to test the long-time hybrid dynamics, we have considered a rational correction term,  which is tailored to grow asymptotically as $\sim q^2$. This is discussed in the next section.

\subsection{Rational correction terms}

\begin{figure}[t]
\center
\includegraphics[scale=.3]{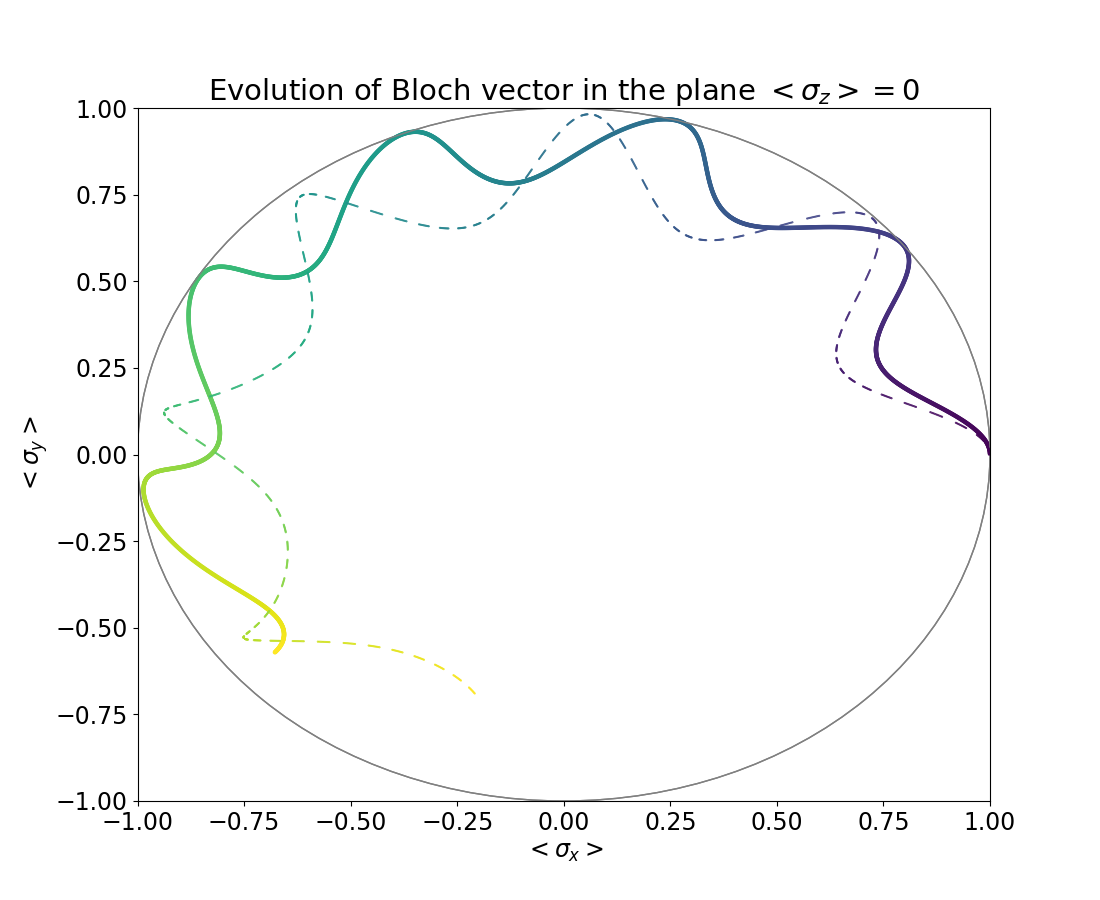}
\includegraphics[scale=.3]{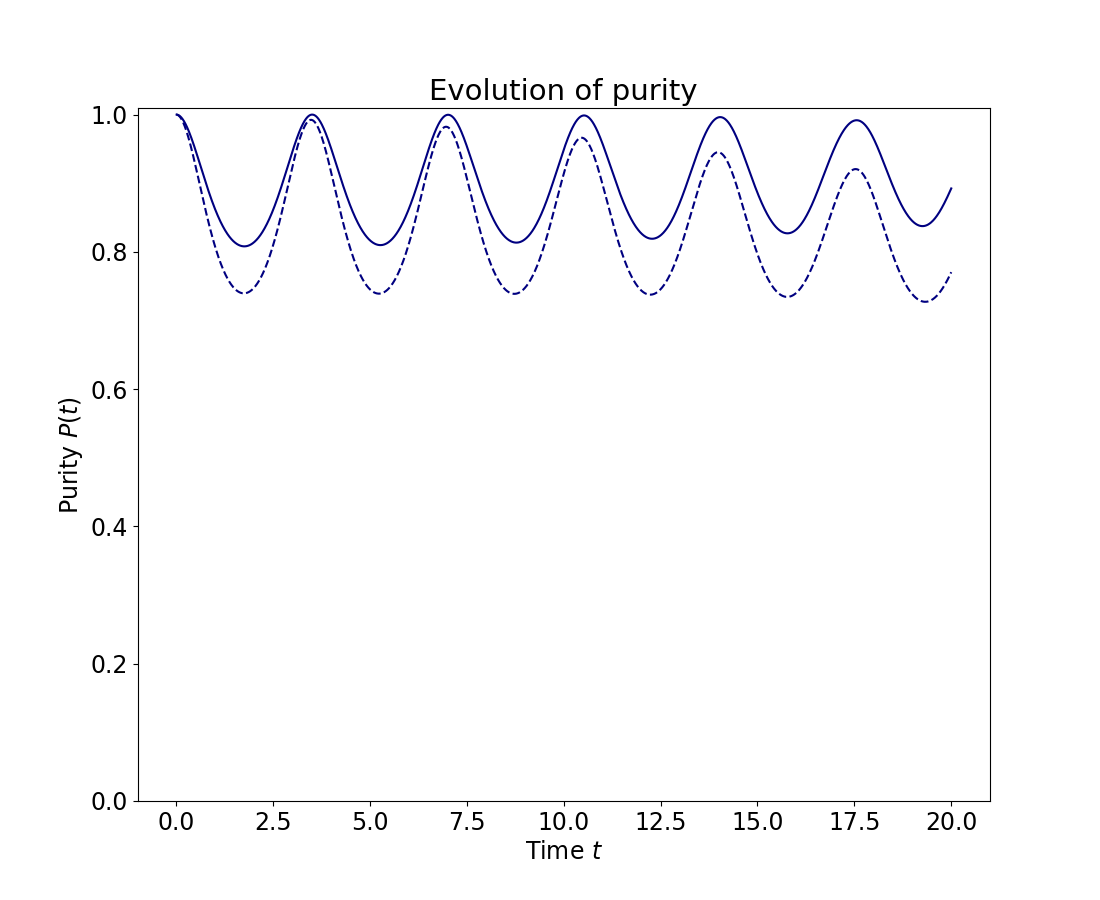}
\includegraphics[scale=.3]{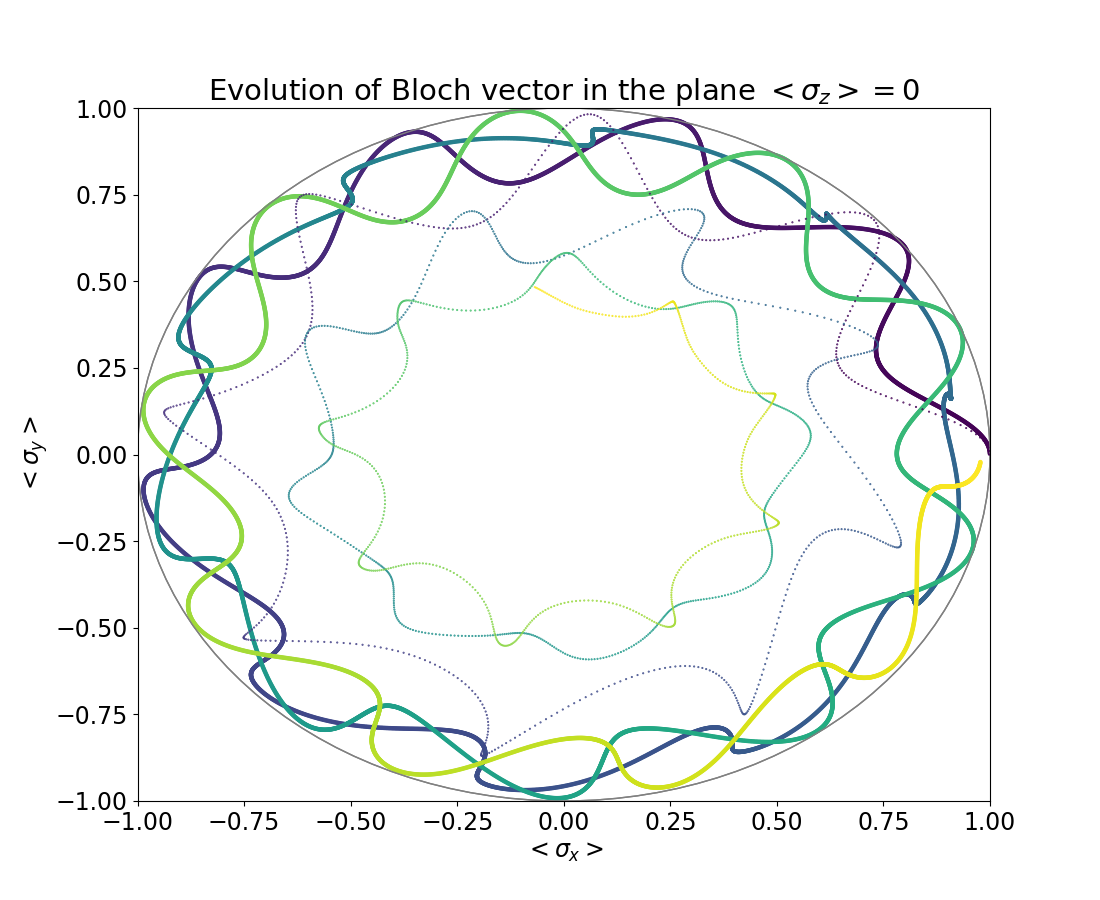}
\includegraphics[scale=.3]{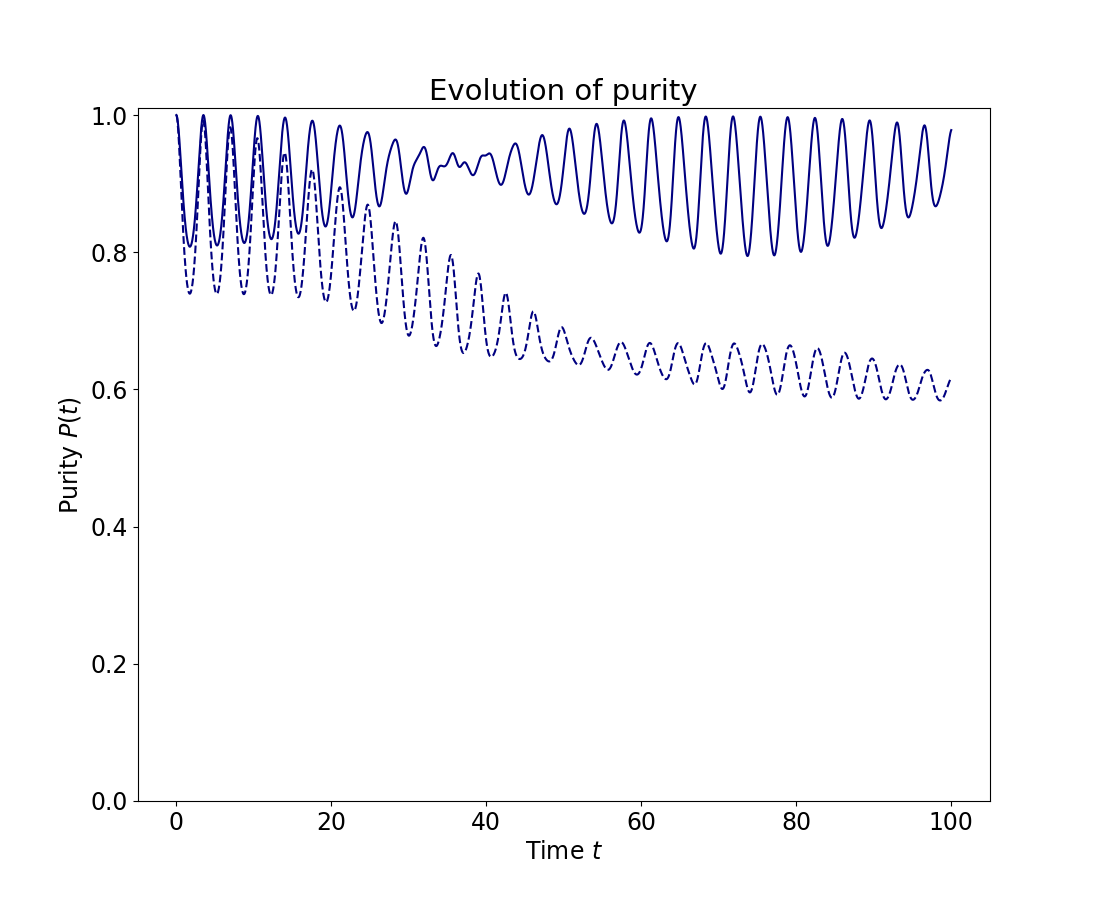}
\caption{{\it Dynamics of the quantum subsystem (rational potential correction)}. Bloch and purity dynamics (left and right panels, respectively) corresponding to the pure-dephasing dynamics for a harmonic oscillator with a rational potential correction. The top panels illustrate the short-time dynamics, while the bottom panels represent the long-time motion.
}
\label{MitraPot}
\end{figure}

In this case, we modify the dynamics from Section \ref{sec:quadcoup} by adding a rational correction  term \cite{carinena,Mitra,RoJaPr08}
$  \alpha q^4/(1+ 2\alpha q^2)$  to either the classical Hamiltonian $H_C$ or the coupling $H_I$.
Hence, we consider hybrid Hamiltonians \eqref{PDHam} with either
\beq
H_C(p,q) = \frac{p^2+\omega^2 q^2}{2}+ \epsilon\, \frac{\alpha q^4}{1+ 2\alpha q^2}
\,,\qquad\quad
H_I(p,q) = {\eta \over 2}(q^2-p^2),
\eeq
or
\beq
H_C(p,q) = \frac{p^2+\omega^2 q^2}{2}
\,,\qquad\quad
H_I(p,q) =   {\eta \over 2}(q^2-p^2) +\epsilon_c\, \frac{\alpha q^4}{1+ 2\alpha q^2}.
\eeq
For this set of simulations, we use the following parameters: oscillator frequency $\omega=1$, quadratic coupling $\eta=0.5$, external field $B_0=0.1$, and inverse temperature $\beta=2$. In addition, we set $\alpha=0.1$.
Compared to the quartic correction of the preceding section, which dominates at large values of $q$, the rational potential is strongly anharmonic at small values of $q$ while it becomes again quadratic for $q \to \infty$.

\begin{figure}[t]
\center
\includegraphics[scale=.3]{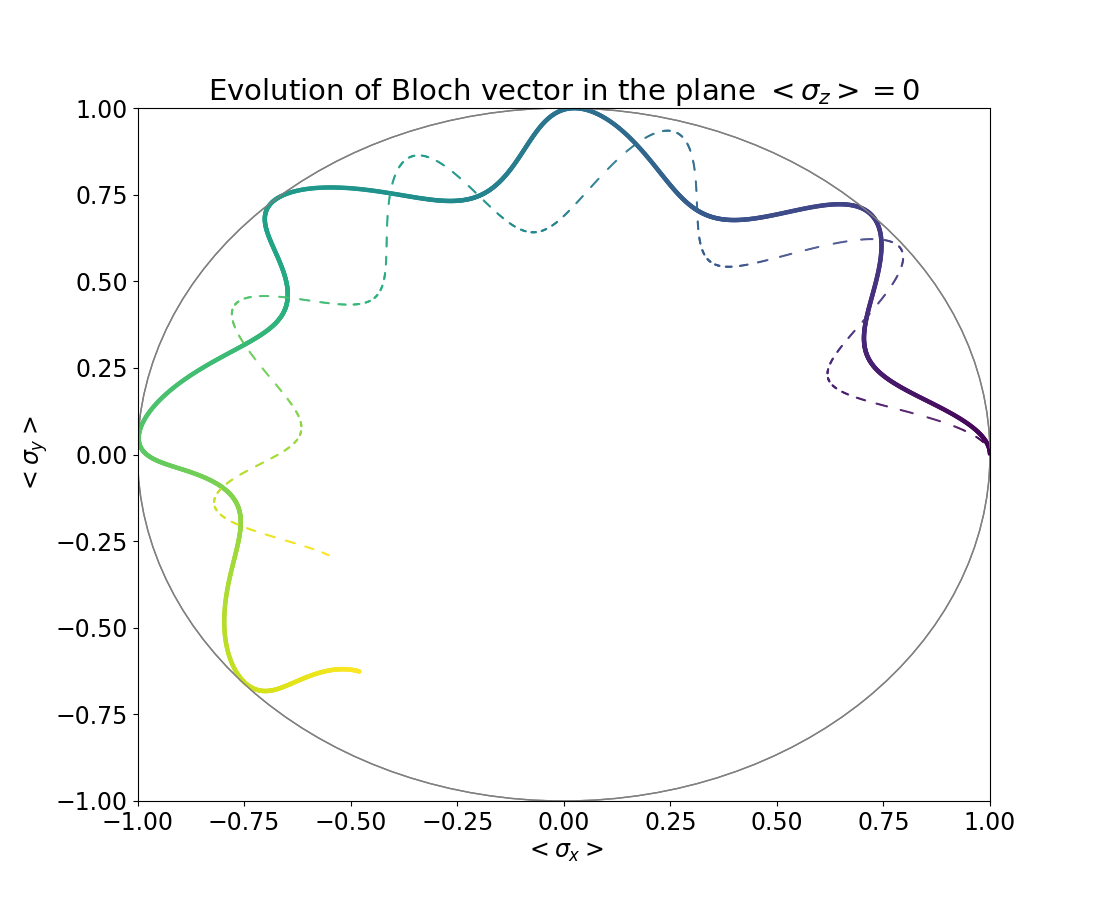}
\includegraphics[scale=.3]{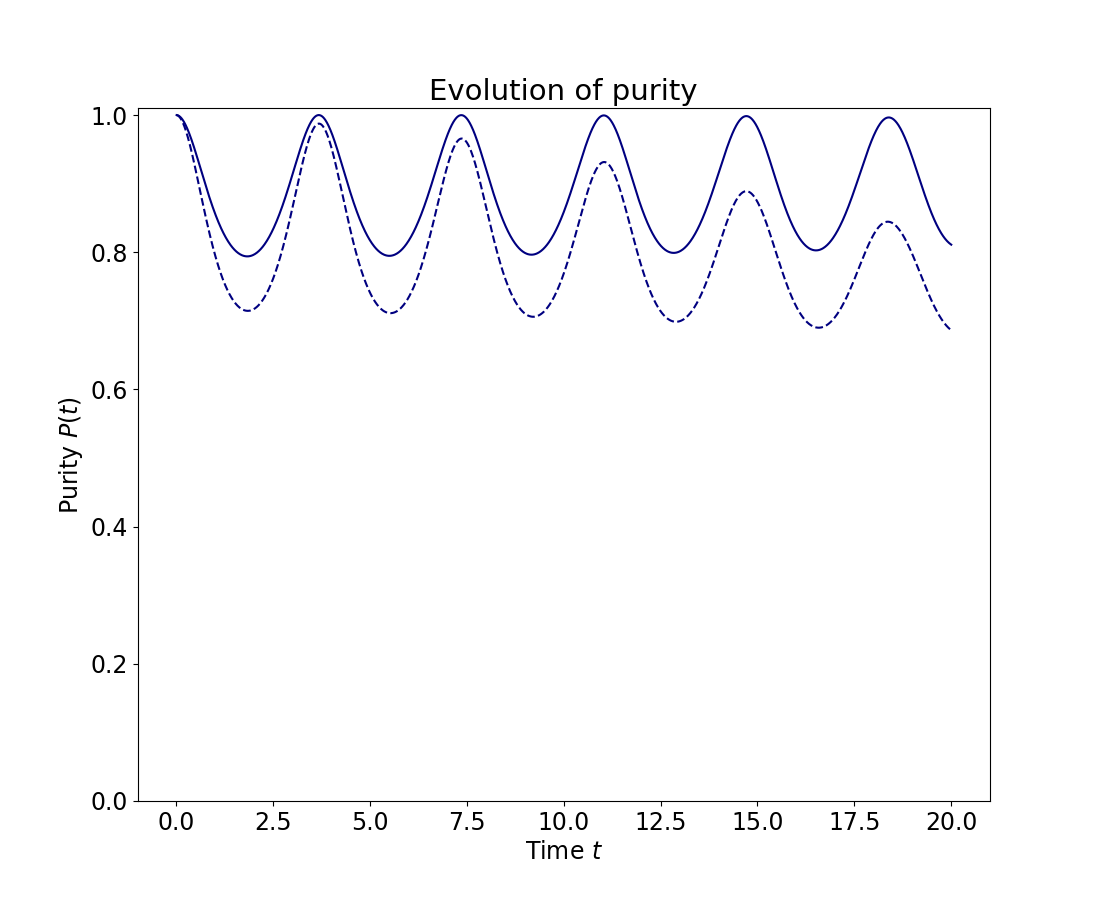}
\includegraphics[scale=.3]{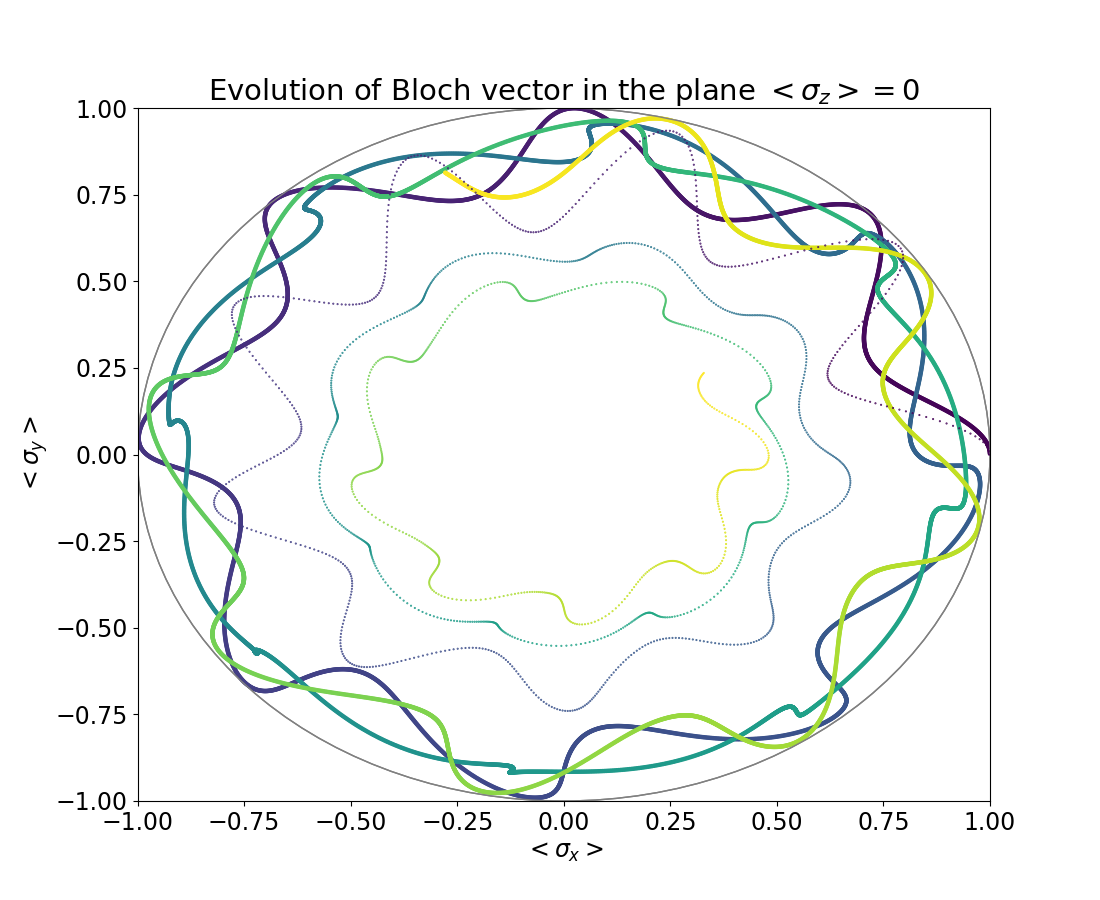}
\includegraphics[scale=.3]{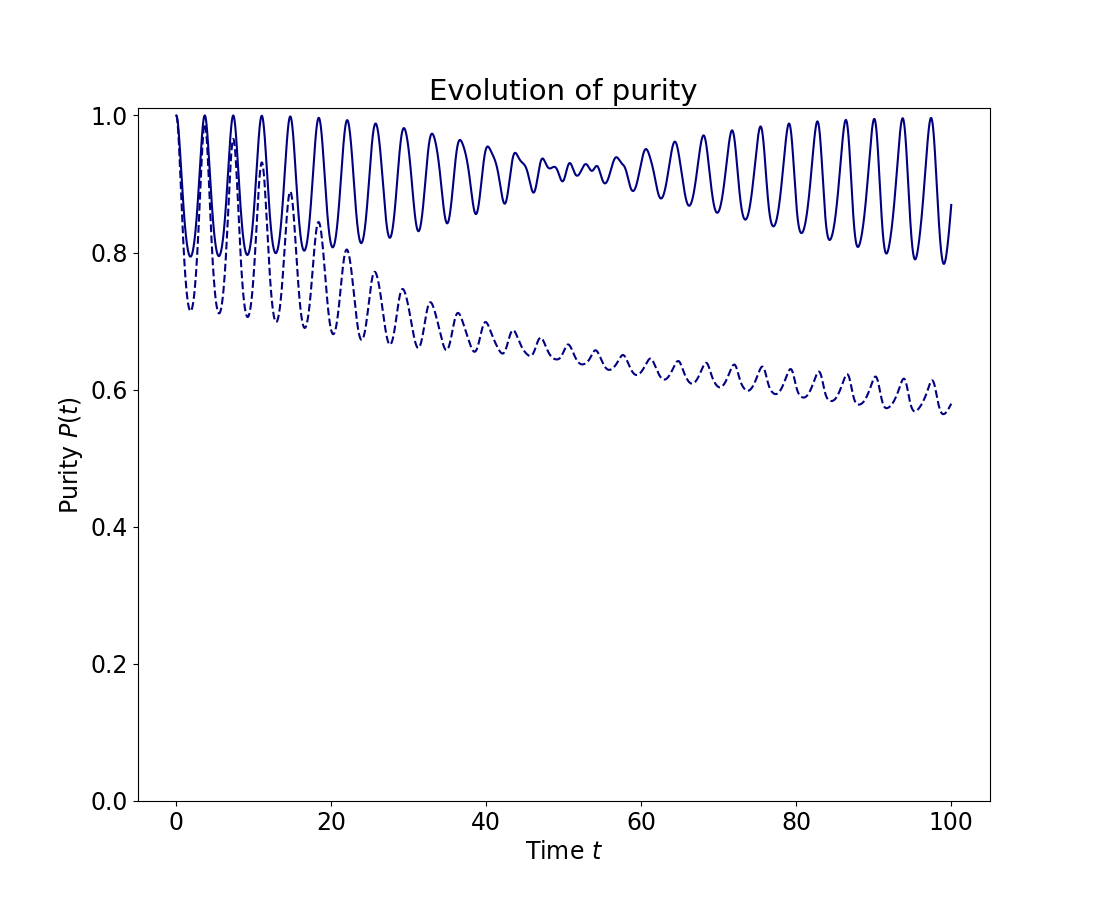}
\caption{{\it Dynamics of the quantum subsystem (rational coupling correction)}. Bloch and purity dynamics (left and right panel, respectively) corresponding to the pure-dephasing dynamics for a harmonic oscillator with a rational coupling correction. The top panels illustrate the short-time dynamics, while the bottom panels represent the long-time motion.
}
\label{MitraCoupl}
\end{figure}

In the first case, we consider a correction of the classical potential ($\epsilon=0.1$ and $\epsilon_c=0$), while in the second case we insert a correction in the coupling  ($\epsilon=0$ and $\epsilon_c=0.05$).
The results are plotted respectively in Fig. \ref{MitraPot} and Fig. \ref{MitraCoupl}, for short times ($t \in [0,20]$, top panels) and long times ($t \in [0,100]$, bottom panels).
For both cases, the short-time hybrid dynamics reproduces relatively well its fully quantum counterpart. Nevertheless, we note a significant difference: for a correction in the classical potential, the Bloch rotation appears faster in the hybrid dynamics than in the quantum evolution, whereas the opposite occurs when the correction applies to the coupling term. This is similar to what was found in the case of a quartic correction.
\begin{figure}[h!]
\center
\includegraphics[scale=.3]{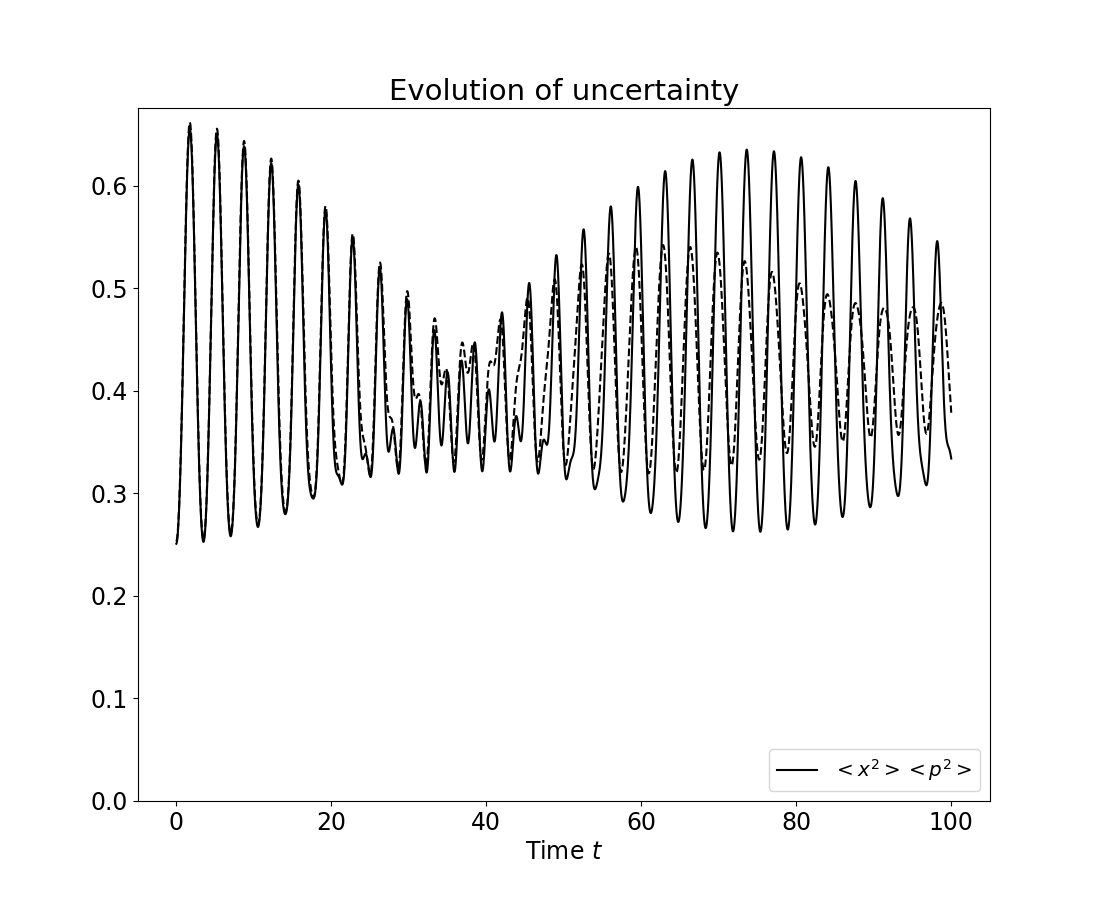}
\includegraphics[scale=.3]{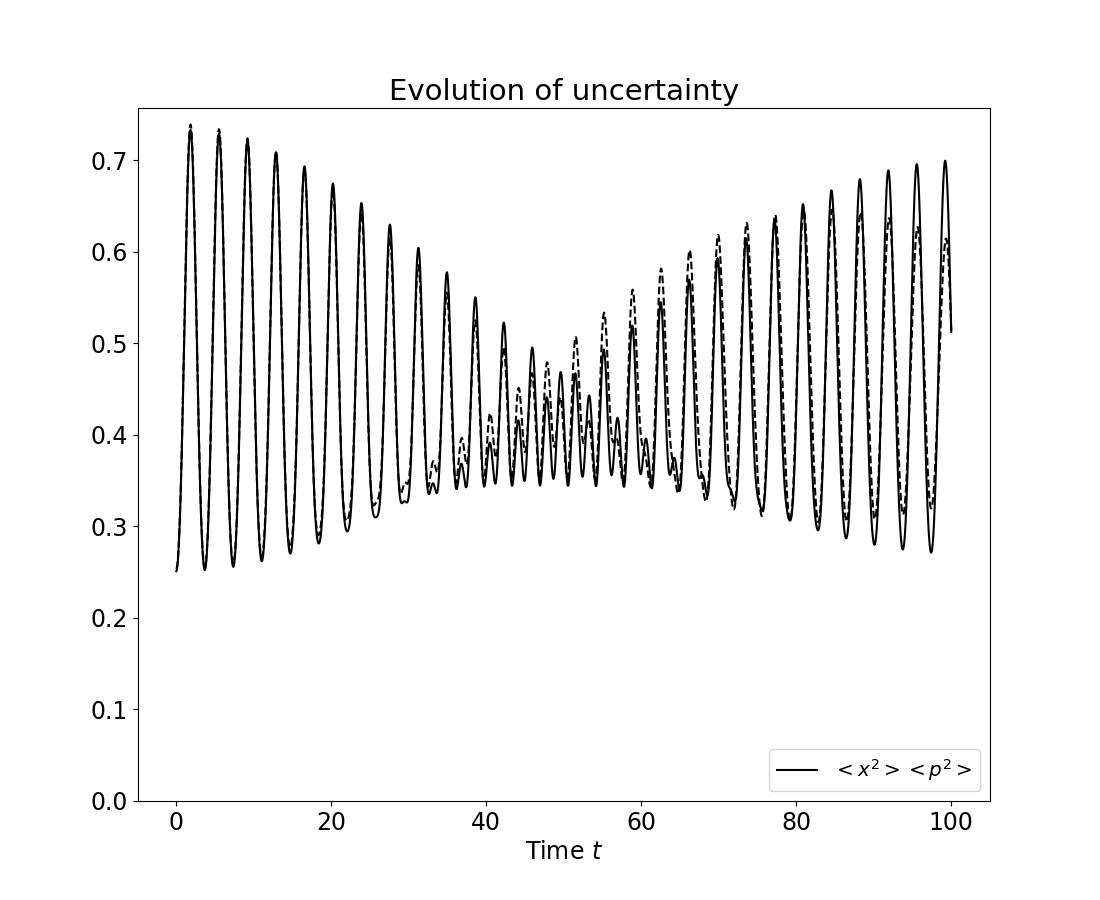}
\caption{ {\it Uncertainty evolution (rational potential/coupling correction)}. Evolution of the uncertainty $\langle q^2\rangle\langle p^2\rangle$ for the case of rational potential correction (left)  and rational coupling correction  (right). Once again, solid lines correspond to the fully quantum result, while dashed lines identify the hybrid dynamics.
}
\label{Uncertainty}
\end{figure}
At longer times, the quantum dynamics found by the hybrid model deviates from the fully quantum predictions. The average value of purity appears to decay again, although this decay is slower compared to the case of a corrected classical potential.

Finally, we also present a comparison of the uncertainty evolution in the quantum and hybrid cases; see Fig. \ref{Uncertainty}. For short times, the hybrid predictions are in excellent agreement with the fully quantum results. For long times, the behaviour is very different depending on whether the rational correction term is inserted in the potential or in the coupling. In the first case, we observe uncertainty oscillations with a much smaller amplitude in the case of the hybrid dynamics, while in the second case this difference in amplitude is less pronounced so that the hybrid model still yields a certain agreement with the fully quantum dynamics.

\section{Conclusions}
A fully quantum description of complex  systems is often a fierce challenge, both for analytical developments and numerical calculations, especially for objects containing as many as hundreds or thousands of particles. In those cases, it may be necessary to separate the system under study into various subsystems, some of which are treated classically and some quantum mechanically. For instance, we may consider a quantum object immersed in a solution of water molecules that are described classically.

However, quantum and classical motion are governed by very different mathematical descriptions, i.e.,  point trajectories in the phase space for  classical systems and wave equations in a Hilbert space for the quantum dynamics. Hence, coupling a quantum system to a classical one is by no means an easy task.
Several approaches to couple quantum and classical systems have been proposed in the past, but each of them presents some drawbacks. Pure-dephasing Hamiltonians are an especially challenging test case for hybrid quantum-classical models and popular approaches  are known to struggle. An example is given by the Ehrenfest model, in which the classical system  decouples completely from the quantum degrees of freedom.

Here, we have used pure-dephasing systems as a benchmark test ground for a recently developed method, jointly proposed by one of us \cite{BoGBTr19}. The method is based on the theory of Koopman wavefunctions, which is a representation of classical mechanics based on the same Hilbert-space formalism as in quantum mechanics (self-adjoint operators, unitary evolution, etc\dots). This feature makes it  a compelling candidate to represent hybrid systems that are partly quantum and partly classical. The resulting quantum-classical wave equation (QCWE) governs the dynamics of a hybrid wavefunction comprising both quantum and classical properties.

Our cases of study  focussed  on the hybrid dynamics of a classical oscillator coupled to a quantum two-level system. In the fully quantum case involving a harmonic oscillator, these test cases emerge as limit cases of the Jaynes-Cummings model  \cite{BlEtAl04} (large-detuning approximation) or, more generally, Rabi models. See also \cite{ReSiSu96} for the experimental relevance of this type of dynamics in the context of electron-phonon coupling at low temperatures.
The hybrid  Koopman equations were implemented into a numerical code that follows the evolution of the hybrid wave function (or rather, its amplitude and phase) in the classical phase space. The long tail of the initial condition \eqref{ICexp} requires special care in the numerical implementation.

For several Hamiltonians involving a harmonic oscillator with linear or quadratic coupling, an overall good agreement was found between the QCWE model and the fully quantum description. For example, the classical  evolution coincides exactly with the oscillator dynamics predicted by the fully quantum treatment. Also, the Bloch sphere motion and the purity evolution (quantifying quantum decoherence)  appear to be well reproduced by the hybrid model. This good agreement persists in the case of a time-dependent coupling, while the case of a time-varying classical frequency leads to very peculiar Bloch rotation dynamics in the fully quantum description, and this feature  escapes from the quantum-classical treatment.

For non-quadratic Hamiltonians, the quantum and hybrid dynamics agreed sufficiently well for short times. For long times, the quantum purity tends to decrease in the hybrid description, while its average levels appear unvaried in the fully quantum treatment. This means that we observed a higher level of decoherence in the hybrid description. Also, unlike the predictions of the Ehrenfest model, the QCWE retains the quantum backreaction force on the classical evolution in all the considered cases.

The present results suggest that the  Koopman approach represents a  valuable step forward in modeling hybrid quantum-classical dynamics. Indeed, while overcoming several outstanding challenges, this approach has recently led to an upgrade model that, unlike the QCWE  used here, is successful in retaining the positivity of the classical Liouville density while treating classical phases as unmeasurable quantities \cite{GBTr23,GBTr22,GBTr21}. Current computational efforts are focusing also on this more advanced approach. For now, we emphasize that the present work is  a \emph{proof-of-principle} study of the QCWE mathematical model and we do not dwell upon the computational complexity of our numerical implementation. On the one hand, the presence of characteristic curves for the evolution of the density $D=\|\Upsilon\|^2$ seems to offer some advantages over fully quantum treatments and certain quantum-classical descriptions \cite{Kapral}. On the other hand, the long tail in the initial condition \eqref{ICexp} may pose new challenges, e.g. involving the statistical sampling. Nevertheless, we believe that the QCWE and its recent nonlinear upgrade  provide a valuable platform for the identification of new convenient multiscale tools in computational physics and chemistry. In the context of nonadiabatic molecular dynamics, for example, ``the parallel development of rigorous but computationally expensive methods and more approximate but computationally efficient methods'' is regarded as absolutely ``critical'', as discussed in \cite{HaSc22}. Other questions also involve the level of applicability of the quantum-classical treatment itself, which may or may not always be a good approximation depending on the fully quantum problem under study. It appears that little is known about this particular problem beyond the standard literature on Born-Oppenheimer molecular dynamics in the adiabatic regime.

Future work will consider more realistic models, beyond the pure-dephasing Hamiltonians studied here. It will also be interesting to consider more complex classical subsystems, e.g., comprising many interacting classical particles.

\medskip
\paragraph{Acknowledgments.} This work arises from our discussions during the conference \emph{Koopman Methods in Classical and Classical-Quantum Mechanics} held online on 18-23 April 2021, during the times of COVID-19. We are  grateful to the Organizing Committee and in particular to the Wilhelm und Else Heraeus Foundation at Bad Honnef, Germany, for  the support provided and for hosting the meeting on their online platform. Also, we are grateful to Paul Bergold, Denys Bondar, Francesco Di Maiolo,  Darryl Holm, and Raymond Kapral for their keen remarks on these and related results.  GM and CT acknowledge financial support from the Royal Society Grant IES\textbackslash R3\textbackslash203005. These results were also made possible through the support of Grant 62210 from the John Templeton Foundation. The opinions expressed in this publication are those of the authors and do not necessarily reflect the views of the John Templeton Foundation.

\bigskip
\addtocontents{toc}{\protect\setcounter{tocdepth}{0}}
\appendix

\section{Numerical solution of the phase-space equations}\label{app:vlasov}
The transport equations \eqref{eq:vlasov1}-\eqref{eq:vlasov2} for the hybrid model both take the following form:
\begin{equation}
{\partial_t f}
+V(p){\partial_q f}+U(q){\partial_p f} - \mathcal{L}(q,p) = 0\, ,
\label{eq:vlasov}
\end{equation}
where $V(p)=\partial_p H$ and $U(q)=-\partial_q H$. These are Vlasov equations, well-known in the plasma physics community.
They can be solved by projecting the phase-space density $f(q,p,t)$ on a phase-space
grid $f_{ij}(t) = f(q_i, p_j, t)$ and using a split-operator method for the time stepping.

The split-operator technique amounts to solving in sequence the following equations:
\begin{eqnarray}
{\partial_t f} +V(p){\partial_q f} &=& 0, \,\,\,\, {\rm for} \,\,\,\Delta t/2 \label{eq:A2}\\
{\partial_t f} +U(q){\partial_p f}  &=& 0, \,\,\,\, {\rm for}\,\,\, \Delta t/2 \label{eq:A3}\\
{\partial_t f} - \mathcal{L}(q,p) &=& 0, \,\,\,\, {\rm for} \,\,\Delta t  \label{eq:A4}\\
{\partial_t f} +U(q){\partial_p f}  &=& 0, \,\,\,\, {\rm for}\,\,\, \Delta t/2 \label{eq:A5}\\
{\partial_t f} +V(p){\partial_q f}  &=& 0, \,\,\,\, {\rm for} \,\,\,\Delta t/2 . \label{eq:A6}
\end{eqnarray}
Note that the above scheme is time-symmetric, and hence exact to second order in the time step $\Delta t$.
Each of the above equations possesses an exact solution. For instance, for Eq. \eqref{eq:A2}, one has:
\beq
f(q,p,t+\Delta t/2) = f(q-V(p)\Delta t/2,p,t) , \label{eq:shift}
\eeq
i.e. a shift of $V(p)\Delta t/2$ in the $q$ direction. Similarly, for Eq. \eqref{eq:A3} one gets a shift of $U(q)\Delta t/2$ in the $p$ direction. The solution of Eq. \eqref{eq:A4} is trivial.

Although the shifts \eqref{eq:shift} are exact, they will not generally map one grid point onto another, hence the need for an interpolation technique. Several methods have been used in the past, ranging from cubic splines to spectral methods. Here, we have used a finite-volume scheme, which has the advantage of locally conserving the transported quantity exactly \cite{Filbet2001}.
For most simulations reported in this work, we used a computational box $[-25, 25] \times [-25, 25]$ (position space $\times$ momentum space), with  $N_q=1000$ points in position space and $N_p=1000$ points in momentum space. The time step was typically $\Delta t = 0.01$.

Grid-based codes generally display good accuracy properties, as they mesh the entire phase space with the same resolution, so that even regions of low phase-space density are well sampled. This is in contrast with trajectory-based methods which sample the phase-space density using pointlike marker ``particles'': in regions where the density is low the number of markers is also low, leading to poor statistical sampling.
In addition, since each of the steps \eqref{eq:A2}-\eqref{eq:A6} has an exact solution in time, grid-based methods do not suffer from limitations due to the Courant-Friedrichs-Lewy (CFL) conditions.

The main limitation of grid-based codes applied to Vlasov-like equations such as \eqref{eq:vlasov} is their significant computational cost, due to the necessity of meshing the entire phase space, which can become a hindrance for high-dimensional configurations. In the present context, this issue is even more compelling because the initial condition \eqref{ICexp} decays algebraically instead of exponentially, as is commonly the case for initial conditions described by a Maxwell-Boltzmann density.

\section{Solution of the Schr\"odinger equations}\label{app:schrodinger}

If the Hamiltonian is at most quadratic in the phase space variables $(q,p)$, then it is possible to find a semi-analytical solution to the Schr\"odinger equation:
\beq \label{eq:schrod}
i \partial_t{\psi_\pm} = H_\pm \psi_\pm
\eeq
for the spinor $\psi = (\psi_+, \psi_-)^T$. For the case of quadratic coupling considered in the main text, the Hamiltonian is:
\begin{equation}
 H_\pm = \frac{\omega^2\left(t\right)q^2+p^2}{2}\pm\frac{\eta(t)}{2}\left(q^2-p^2\right)\pm B_0\left(t\right) .
\end{equation}

It can be shown that an exact solution to the time-dependent Schr\"odinger equation takes the Gaussian form
\beq
\label{eq:gaussianwv}
\psi_\pm\left(q,t\right)=A_\pm\left(\frac{2b_\pm}{\pi}\right)^\frac{1}{4}
\exp\!\Big({-b_\pm\left(q-d_\pm\right)^2}+{ia_\pm\left(q-d_\pm\right)^2+ie_\pm\left(q-d_\pm\right)+ic_\pm\Big)} ,
\eeq
provided the initial condition is also in this form. In the above expression, the auxiliary functions  $a_\pm(t)$, $b_\pm(t)$, $c_\pm(t)$, $d_\pm(t)$ and $e_\pm(t)$ are real functions that depend on the time $t$, and $A_\pm$ are normalization constants chosen so that $\left|A_+\right|^2+\left|A_-\right|^2=1$.

By injecting the Gaussian wave function \eqref{eq:gaussianwv} into the Schr\"odinger equation \eqref{eq:schrod} and equating the coefficients of the terms  in $q^0$, $q^1$ and $q^2$, one obtains that the above auxiliary functions should  obey the following set of ordinary differential equations:
\begin{equation}
\begin{cases}
\dot{a_\pm}=-\frac{\omega^2\left(t\right)\pm\eta(t)}{2}-2\left(a_\pm^2-b_\pm^2\right)[1\mp\eta(t)] ,
\\\dot{b_\pm}=-4a_\pm b_\pm [1\mp\eta(t)] ,
\\\dot{c_\pm}=-\frac{\omega^2\left(t\right)\pm\eta(t)}{2}\,d_\pm^2+\left(\frac{e_\pm^2}{2}-b_\pm\right)\left[1\mp\eta(t)\right]\mp B_0\left(t\right) ,
\\\dot{d_\pm}=e_\pm [1\mp\eta(t)] ,
\\\dot{e_\pm}=-d_\pm [\omega^2(t)\pm\eta(t)] ,
\end{cases}
\label{eq:ODE}
\end{equation}
where a dot stands for differentiation with respect to time. As initial condition at $t=0$, we take:  $b_\pm (0)=1/2$ (minimum uncertainty wave packet), $d_\pm(0)=q_0$  and $e_\pm (0)=0$ (Gaussian wave packet centered at $q_0$), and $a_\pm (0)=c_\pm (0)=0$ (vanishing initial phase).

The equations \eqref{eq:ODE} form a system of nonlinear ODEs, which can be solved numerically with little effort, using for instance a standard fourth-order Runge-Kutta scheme. Once the auxiliary functions are known (numerically) for all times, it is easy to reconstruct any observable from the wave function \eqref{eq:gaussianwv} at time $t$.
For instance, the mean position is:
\[
\langle q \rangle = d_+\left|A_+\right|^2+d_-\left|A_-\right|^2 ,
\]
and the $z$ component of the spin
\[
\langle{\sigma_z}\rangle =\int{\left|\psi_{+}\right|^2\ dq}-\int{\left|\psi_{-}\right|^2\ dq}=\left|A_+\right|^2-\left|A_-\right|^2 ,
\]
which also shows that $\langle{\sigma_z}\rangle$ is a constant of the motion, as expected.
For other variables, the expressions are more convoluted, but can still be easily computed.

For Hamiltonians other than quadratic, the above procedure cannot be used and a fully numerical solution of the  Schr\"odinger equation \eqref{eq:schrod} has to be envisaged. For the anharmonic Hamiltonians considered here, we chose a standard Crank-Nicolson scheme, which is second-order accurate in time and space.

\end{document}